\documentclass[prb,twocolumn,amsmath,amssymb,10pt,aps,unsortedaddress, superscriptaddress]{revtex4-1} 

\usepackage{soul}
\usepackage{amsmath}
\usepackage{graphicx}
\usepackage{rotating}
\usepackage{multirow}
\usepackage{latexsym}
\usepackage{textcomp}
\usepackage{verbatim}
\usepackage{color}
\usepackage{pict2e}
\usepackage{bm}
\usepackage{subfigure}
\usepackage{everysel}
\usepackage{keyval}
\usepackage{ragged2e}
\usepackage{amssymb}
\usepackage{enumitem}
\usepackage{pstricks}
\usepackage{mathtools}
\usepackage{hyperref}
\usepackage{braket}
\usepackage{slashed}
\usepackage{mathrsfs}

\newcommand{\osum}{{%
    \setbox0\hbox{\circ}%
    \rlap{\hbox to \wd0{\hss\sum\hss}}\box0
}}

\begin{document}

\title{Anomaly Manifestation of Lieb-Schultz-Mattis Theorem and Topological Phases}   % type title between braces

\author{Gil Young Cho}         % type author(s) between braces
\thanks{The first two authors contributed equally to the work.}
\affiliation{School of Physics, Korea Institute for Advanced Study, Seoul 02455, Korea}
\affiliation{Department of Physics, Korea Advanced Institute of Science and Technology, Daejeon 305-701, Korea}

\author{Chang-Tse Hsieh}
\thanks{The first two authors contributed equally to the work.}
\affiliation{Department of Physics, University of Illinois, 1110 W. Green St., Urbana, Illinois 61801-3080, USA}

\author{Shinsei Ryu}
%\affiliation{Department of Physics, University of Illinois, 1110 W. Green St., Urbana, Illinois 61801-3080, USA}
\affiliation{James Franck Institute and Kadanoff Center for Theoretical Physics, University of Chicago, Illinois 60637, USA}
\date{\today}    % type date between braces

\begin{abstract} 
 % The Lieb-Schultz-Mattis theorem constrains
 % the possible low-energy and long-distance behaviors of 
 % states which emerge from microscopic lattice Hamiltonians.
  The Lieb-Schultz-Mattis (LSM) theorem dictates
  that emergent low-energy states from a lattice model
  cannot be a trivial symmetric insulator
  if the filling per unit cell is not integral
  and if the lattice translation symmetry and particle number conservation
  are strictly imposed.
  In this paper, we compare the one-dimensional gapless states
  enforced by the LSM theorem 
  and the boundaries of one-higher dimensional strong symmetry-protected
  topological (SPT) phases from the perspective of quantum anomalies. 
  We first note that,
  they can be both described by the same low-energy effective field theory
  with the same effective symmetry realizations on low-energy modes,
  %and low-energy spectrum, 
  %the gapless states emergent from lattice Hamiltonians are equivalent to the
  %boundary theory of the SPT state,
  wherein non-on-site lattice translation symmetry is encoded as if it is a local symmetry.
  %Once a global symmetry is realized in a non-on-site fashion, 
 % Hence, the boundary of the topological phases can be realized in a stand-alone lattice model,
  %Hence, the no-go theorem for the boundary is circumvented
  %in a similar fashion as the recent discussions of the half-filled Landau level and  topological insulators.
  In spite of the identical form of the low-energy effective field theories,
  %Though the edge of the SPT states and the lattice systems have the identical
  %effective theories,
  we show that the quantum anomalies of the theories
  play different roles in the two systems.
  In particular, we find that the chiral anomaly is equivalent to the LSM
  theorem, whereas
  there is another anomaly,
  which is not related to the LSM theorem but is intrinsic to the SPT states.
  % Using these understandings of the theorem,
  As an application,
  we extend the conventional LSM theorem to multiple-charge
  multiple-species problems
  and construct several exotic symmetric insulators.
  We also find that the (3+1)d chiral anomaly provides only
  the perturbative stability of the gapless-ness local in the parameter space.

\end{abstract}

\maketitle

\tableofcontents

\section{Introduction}

%\subsection{Introduction}

Predicting possible macroscopic behaviors of many-body systems 
from a given kinematical input data,
such as spatial dimensions, the presence of a certain set of symmetries, etc.,
is a central question in many-body physics. 
More precisely, predicting spectral properties 
(e.g., presence/absence of a spectral gap above ground states)
and the nature of ground states (e.g., long/short-range entangled, trivial, etc.) 
would be of great interest. 

In this regard, we will discuss the following three classes of problems (systems) in this paper:
%in which such prediction is possible.  

\begin{itemize}
\item[(i)] {\it The LSMOH theorem}:
The Lieb-Schultz-Mattis theorem and its generalization by Oshikawa and Hastings 
\cite{Lieb1961, Oshikawa2000, Hastings2004, Chen2011}
dictates that when the lattice translation symmetry and $U(1)$ charge (electric charge, spin, etc.) conservation
are preserved, the system must be gapless or its ground state must be long-range entangled
if the particle number (or spin quantum number) per unit cell is fractional (non-integral). 
In one spatial dimension, this in particular means that the system has to be gapless.

\item[(ii)] {\it SPT boundaries}: 
The boundaries of a symmetry-protected topological (SPT) phase\cite{KaneRev, QiRev,SenthilReview}
cannot be gapped trivially, i.e., 
they must be either gapless or exhibit topological order,  
so far as the symmetries protecting the bulk SPT phase are enforced.
For (1+1)-dimensional boundaries of (2+1)-dimensional SPT phases, 
this in particular means that they have to be gapless.

\item[(iii)] {\it Fermi ``surfaces''}:  
There are a class of lattice fermion systems in which 
the single-particle spectrum supports zeros in the Brillouin zone,
i.e., Fermi (nodal) points/lines/surfaces, etc.,  
%are at least {\it perturbatively} stable,
in the presence of a certain set of symmetries. 
For notational simplicity, we will call such zeros of the single particle spectrum 
Fermi ``surfaces'', although one should bear in mind that such zeros can form a hypersurface of various dimensions.
%do not have to form a surface (but can be either point or line). 
\end{itemize}

%We note that an important distinction between
We distinguish systems in Class (iii) from those in
the other classes
by their {\it perturbative} stability.
I.e., the gapless nature in Class (iii) is only {\it perturbatively} or
{\it locally} stable;
The impossibility of trivial gapped states dictated by the LSMOH theorem
and at SPT boundaries is {\it non-perturbative} in the sense that it relies
only on the kinematical input data (e.g., the filling faction, symmetries),
but not on the interaction strength. 
On the other hand, the local stability in Class (iii) excludes, in the parameter space, 
symmetric trivial insulators only in the vicinity of a given gapless low-energy theory.
In other words, a trivial insulator may exist
if the system is perturbed far away from the low-energy theory. 
This may become particularly important when the Fermi ``volume'' inside the surface is zero, e.g., nodal points and lines.
One way to understand the perturbative stability of Fermi surfaces is to note that 
since by the fermion-doubling theorem, these Fermi surfaces always appear in pair.
Hence unless enough symmetry conditions are imposed, by adding strong enough perturbations, 
these systems are ultimately gappable by ``pair annihilating'' these Fermi surfaces. 
Nevertheless, some of Fermi surfaces are expected to be stable locally or perturbatively.
With further symmetry constraints, it would be possible to turn systems in Class (iii) into 
systems in Class (i), which are stable beyond the perturbative level.
For example, for the case of Fermi surfaces with a finite Fermi volume in Class (i),
imposing translation and charge conservation symmetries turns the system into the class (i).

%In this paper, we will distinguish the “local stability”
%(which is our shortened version of local-in-parameter-
%space stability) and “non-perturbative stability”.
%The non-perturbative stability of the gapless-ness relies only on the microscopic data, filling and symmetry.
%According to the LSM theorem, the 1d fractionally-filled
%lattice model must be gapless if the translation symmetry and charge
%conservation are imposed.
%In different words, there is no symmetric trivial insulator in the
%whole parameter space. Hence the gapless-ness is non-
%perturbatively stable.
%On the other hand, the local stability only excludes the
%symmetric trivial insulator in the local parameter space
%around the given low-energy theory. There may be a trivial insulator, which may be far away from the given low-
%energy theory.

Among (i-iii), the ``mechanism'' behind the obstruction for trivially gapping out the
SPT boundaries [class (ii)] is understood in terms of quantum anomalies.
\cite{Cho2014,Hsieh2014,Hsieh2016,Cho2015, Witten2015, Chen2015,ryu2012electromagnetic, Kapustin}
In particular, when the symmetries protecting SPT phases are unitary and 
strictly local (``on-site''), i.e., for ``strong'' SPT phases, 
%(this in particular means we do not consider spatial symmetries nor time-reversal symmetry), 
the relevant anomalies are 't Hooft anomalies.
%(The precise meaning of ``on-site'' will be spelled out momentarily.)
Here, a 't Hooft anomaly is an obstruction to gauge on-site global symmetries of the theory.
%Such anomaly can be often manifested
%when the boundary conditions are twisted by (or ``gauging" of) symmetry of SPTs.
%\cite{Levin2012, Ryu2012}

More precisely,
for bosonic systems 
whose Hilbert space $\mathcal{H}$ is factorized into local Hilbert spaces,
$\mathcal{H}= \prod_x \mathcal{H}_x$
where $\mathcal{H}_x$ is the Hilbert spaces for a given ``lattice site'' $x$,
a unitary symmetry $g$ is said to be on-site
if  
%In the typical setting of SPT phases, 
%we start from bulk phases where symmetry actions are strictly local or purely on-site.
$g$ factorizes similarly as $g=\prod_x g_x$.
(This property of $g$ is also called splittable.) 
For fermionic systems, there is no natural factorization of the fermion Fock space due to
the Fermi statistics.
Nevertheless, we will assume that the similar notion of on-site symmetries exists, 
when a unitary operator $g$ transforms
fermion creation/annihilation operators purely locally.

In the typical setting of SPT phases, 
we start from bulk phases where symmetry actions are purely on-site.
At non-trivial SPT boundaries, however,
symmetries cannot be made purely on-site.  
This is in fact another way to state that SPT boundaries suffer from (or enjoy)
quantum anomalies ('t Hooft anomalies). 
%This is in fact closely connected to the non-on-site nature of the symmetry action. 
Boundaries of topologically distinct SPT phases are characterized by
different 't Hooft anomalies.
In fact, the topological invariants characterizing bulk SPT phases
are in one-to-one correspondence with 't Hooft anomalies of SPT boundaries
-- the fact known as the bulk-boundary correspondence.
In other words,
the boundaries of SPT phases cannot exist on their own
(i.e., cannot be put on a proper lattice),
if we require the relevant (unitary) symmetries be strictly local (on-site)
-- the boundary theories of an SPT phase cannot be decoupled or ``disentangled'' from  
its bulk because of the anomalies. 
The impossibility of realizing boundaries of SPT phases as an isolated local system
is usually called as the no-go theorem.

The purpose of this paper is,
by taking simple examples, 
to give a detailed comparison between the LSMOH theorem
and SPT boundaries. 
In particular, given that the impossibility of trivially gapping SPT boundaries is due to quantum anomalies ('t Hooft anomalies),
we will make an attempt to interpret the LSMOH theorem in terms of quantum anomalies. 
For precursors of the current work,  
discussing the relationship between SPT phases and the LSMOH theorem,
see, for example, Refs.\ \onlinecite{furuya2015symmetry, po2017lattice, song2017topological, Cheng2015,thorngren2016gauging}.
We will also touch upon
the origin of the perturbative stabilities of Fermi surfaces using quantum anomalies.
It should be noted that the stability of Fermi surfaces has been so far discussed mainly at 
the level of the single particle physics.
Our discussion using quantum anomalies should shed light on the stability of
Fermi surfaces in the presence of interactions.

In the rest of the Introduction, we will list and briefly describe
some of the key issues in discussing the similarities and distinctions
among the three classes of problems (i-iii). 
See
Sec.\ 
\ref{LSMOH and no-go theorem; On-site v.s. non-on-site symmetries},
\ref{LSMOH v.s. SPT anomalies}
and
\ref{Effective symmetry and perturbative stability}
below.
They also serve as a short summary
for 
Sec.\ \ref{2d SPTs and 1d lattice models}, 
\ref{Anomaly and LSMOH theorem},
and 
\ref{SPT anomaly and Local Stability}
in the main text. 

\subsection{LSMOH and no-go theorem; On-site v.s. non-on-site symmetries}
\label{LSMOH and no-go theorem; On-site v.s. non-on-site symmetries}

To explore a possible connection between the LSMHO theorem and quantum anomalies 
(and SPT boundaries), 
we will first note that
the low-energy physics of
lattice models dictated to be gapless by the LSMOH theorem 
and SPT boundaries
can be described by
an identical continuum field theory.  
For example, in Sec.\ \ref{2d SPTs and 1d lattice models}, 
we will discuss a (1+1)d lattice fermion model at fractional filling.
For the low-energy effective field theory of this model,
we will find that there is a (2+1)d SPT phase
(a version of the quantum spin Hall effect)
whose boundary is described by the same continuum field theory. 
Here, the relevant symmetry
in the (1+1)d lattice fermion model
is the $U(1)$ particle number conservation
and lattice translation symmetry,
whereas on the SPT side
the relevant symmetry is
the $U(1)$ particle number conservation
and
$U(1)$ or $\mathbb{Z}$ internal (spin) 
rotation symmetry.

Since
the LSMOH theorem concerns isolated lattice systems
without referring to any higher dimensional bulk systems,
%that relating the LSMOH theorem to quantum anomalies 
%may look seemingly against the no-go theorem. 
this may look seemingly against the no-go theorem.
%If the impossibility of trivially gapping out dictated by the LSMOH theorem
%were related to quantum anomalies, why would it be possible to realize 
%systems of interest in the context of the LSMOH theorem on an isolated lattice?
The trick of evading the no-go theorem is that, while symmetries in the bulk SPT phases are realized {\it on-site},
symmetries entering into the corresponding LSMOH theorem are {\it non-on-site}. 
In the typical setting of SPT phases, 
we start from bulk phases where symmetry actions are strictly local or
purely on-site.
On the other hand, in the context of the LSMOH theorem, 
it typically involves non-on-site symmetries.
E.g., lattice translation symmetries. 
This is the reason why, even if the LSMOH theorem may be related to some sort of quantum
anomalies, relevant systems can still be put on a lattice without
having a higher dimensional bulk. 
Evading the no-go ``theorem'' is also possible in higher dimensions.
For example, the ``duality'' between the composite Fermi liquid in the half-filled Landau
level and the (2+1)d boundary of (3+1)d topological insulators
has been discussed extensively recently.
\cite{Son2015, Geraedts2016, Wang2015CFL,Metlitski2015,Mross2016,Wang2016half}
See Sec.\ \ref{2d SPTs and 1d lattice models}.

We will also note in Sec.\ \ref{2d SPTs and 1d lattice models}
that the lattice translation symmetry 
within the low-energy field theory 
can be encoded as 
an {\it effective} symmetry.
In the of the rational filling $\nu = p/q$ for mutually-prime $p$ and $q$, 
the translation symmetry (up to some gauge choice and changes in band structures) can be further reduced as $\mathbb{Z}_q$, 
i.e., there may be symmetry-reduction $G = \mathbb{Z} \to G_{eff} = \mathbb{Z}_q $. 
Hence, we may consider the effective translation symmetry $G_{eff} = \mathbb{Z}_q$ as the local symmetry in the low-energy limit.

\subsection{LSMOH v.s. SPT anomalies}
\label{LSMOH v.s. SPT anomalies}
%{\color{red} Need to mention:
%  (i)low-energy theory and
%  (ii) LSMOH anomaly v.s. SPT anomaly
%}

Having confirmed that  
the low-energy theories 
for the fractionally filled 1d lattice fermion model
and
for the SPT boundary are identical,
we will discuss
quantum anomalies within the low-energy theory
in Sec.\ \ref{Anomaly and LSMOH theorem}.
The identification/computation of quantum anomalies can be done
within the low-energy theories
since anomalies are preserved along the RG flow -- the 't Hooft anomaly matching. \cite{Matching} 
If the low-energy effective theory has a 't Hooft anomaly,
one would then
expect that the theory at any energy scale and at any interaction strength
cannot be deformable to a symmetrical trivial insulator.

That an SPT boundary and a lattice model for which we apply the LSMOH theorem 
can be described by the same low-energy effective theory 
would imply that the both systems have the same anomalies.
However, we will show that there are some subtleties --
instead of the full 't Hooft anomaly, we need to consider the chiral anomaly
for the LSMOH theorem.

%In the above discussion,
%we have seen that the setting of the LSMOH theorem is rather similar to SPT boundaries.
%A slight difference between SPT boundaries
%and the LSMOH theorem is that
%in the latter
%the system of our interest is realized in its isolation (without invoking bulk),
%and, as a price, we know that (the part of) the relevant symmetries are explicitly non-on-site.
%In particular, an SPT boundary and a lattice model for which we apply the LSMOH theorem 
%can be described by
%the same low-energy effective theory (continuum field theory).

%As advocated, we will make an attempt to identify quantum anomalies
%for lattice models which are forced to be non-trivial by the LSMOH theorem. 
%%emergent from (1+1)d lattice models and the boundaries of (2+1)d SPT phases.
%%and identify the physical implications of the anomalies in these systems.
%The identification/computation of quantum anomalies can be done
%within the low-energy theories
%since anomalies are preserved along the RG flow.
%(the 't Hooft anomaly matching). 
%If the low-energy effective theory has a 't Hooft anomaly,
%then the theory at any energy scale and at any interaction strength
%cannot be deformable to a symmetrical trivial insulator.

In Sec.\ \ref{Anomaly and LSMOH theorem}
we will illustrate this by considering the (1+1)-dimensional lattice fermion model
at filling $\nu$,
in the presence of the lattice translation symmetry and global $U(1)$ charge
conservation symmetry.  
Here, by the full 't Hooft anomaly,  
we mean the 't Hooft anomaly of the whole global symmetries
(=  an effective on-site version of lattice translation $\times$ charge $U(1)$).
On the other hand,
the chiral anomaly involves the two symmetries
and partially gauging the symmetries, e.g., only one of the two symmetries.
It effectively ``measures'' the conflict of the two
symmetries, or violation of one global symmetry when
the other symmetry is gauged.
This chiral anomaly implies that both the symmetries cannot be gauged
consistently and thus the obstruction to a symmetric trivial insulator.
In some sense, the chiral anomaly can be thought of as
a part of (a subset of) the full 't Hooft anomaly. 

As the chiral anomaly is the subset of the full 't Hooft anomaly
the chiral anomaly gives rise to a ``cruder'' classification of SPT phases
when it comes to the interacting classification.
We will show that the no-go condition for a symmetric insulator
from the LSMHO theorem is identical to the non-trivial chiral anomaly.
The chiral anomaly hence provides a non-perturbative stability of the gaplessness.

%The similar obstruction to a trivial insulator is also
%found in the lattice model subject under the LSM theorem.

%
%Generically, the presence of any of the
%quantum anomalies signals certain amount of the stability
%of the theory against deformation of the theory into
%a symmetric trivial insulator.
%However, the ``degree'' of the anomalies’ protection of the theory against deforming
%into a symmetric trivial gap are different in the two cases.
%For SPT boundaries, any of non-trivial anomalies,
%either chiral or SPT anomalies, implies the obstruction to the symmetric insulator.
%We expect that the stability implied from the anomalies is completely non-perturbative
%and thus the gapless-ness does not depend on the particular realization
%of the edge theory and strength of the
%symmetric perturbations. Note that the chiral anomaly
%is the subset of the ’t Hooft anomaly (since the other
%anomaly, SPT anomaly, may be non-trivial) and hence
%it gives rise ``cruder'' classification when it comes to the interacting classification.
%On the other hand, for the lattice system, the chiral anomaly and the SPT
%anomaly play different roles.
%We show that the no-go condition for a symmetric insulator
%from the LSMHO theorem is identical to the non-trivial
%chiral anomaly. The chiral anomaly hence provides a
%non-perturbative stability of the gapless-ness.

\subsection{Effective symmetry and perturbative stability}
\label{Effective symmetry and perturbative stability}

In contrast to the chiral anomaly,
we will discuss the other part of the full 't Hooft anomaly
(the system with vanishing chiral anomaly)
in Sec.\ \ref{SPT anomaly and Local Stability}.
For the theory emergent from the (1+1)d lattice system,
we will argue that 
the other part of the full 't Hooft anomaly
implies the perturbative stability of Fermi surfaces.
This perturbative stability detected by the anomaly
is the one-dimensional analogue of the classification of (some) nodal fermions
emerging from accidental band crossings in higher dimensions,
e.g., classification of the nodal fermions in (3+1)d systems.\cite{Young2012dirac,yang2014classification,yang2015topological,burkov2011weyl,yang2011quantum,cho2011possible}
This is particularly important when the filling is rational
$\nu= p/q$ (for mutually prime $p$ and $q$).
For the filling, when the band structure is fine-tuned,
the low-energy translation symmetry can be effectively reduced to
$\mathbb{Z}_q$,
a subset of full translation symmetry $\mathbb{Z}$.
Then the anomaly signals that the system must be gapless
only when the translation symmetry is strictly $\mathbb{Z}_q$,
but not bigger than this.
In other words, when the full translation symmetry $\mathbb{Z}$ is considered,
the theory with this anomaly only can be gapped out symmetrically.
However, to have a symmetric gap to the spectrum,
we need non-perturbative processes, e.g., to introduce extra ``trivial''
degrees of freedom, non-quadratic interaction terms or
to change the band structures. 
[Here, the opposite of the perturbatively stable,
i.e., the perturbatively gappable, 
is equivalent to the condition that we can gap out the spectrum within the quadratic terms without any further modification of the given theory. 
This is slightly different from the ``perturbative'' stability
in the renormalization group theory sense, 
i.e., the absence of relevant directions of the theory in the parameter space.
(Note that when the strong forward scattering is present, some multi-fermion terms may become relevant in Luttinger liquids.)]

%through the full band-width scales.
%Hence, the non-trivial SPT anomaly provides the local-in-parameter-space stability for the low-energy theory. 

%For example, when the filling is rational $\nu = p/q$
%(for mutually prime $p$ and $q$),
%the low-energy translation symmetry
%is seemingly $\mathbb{Z}_q$,
%a subset of full translation symmetry $\mathbb{Z}$.
%Then the SPT anomaly signals that the system must be gapless only
%when the translation symmetry is strictly $\mathbb{Z}_q$, but not bigger than this.
%Nevertheless, even when $\mathbb{Z}_q$ is extended to Z, we need to
%introduce extra ``trivial'' degrees of freedom, which is a non-perturbative process,
%to gap out the spectrum symmetrically.

%Hence, the non-trivial SPT anomaly provides the
%local-in-parameter-space stability for the low-energy theory. Note that the presence of the SPT anomaly does
%not preclude the existence of the symmetric insulator in
%the parameter space, but it may be far away from the gapless theory.

The different origins of the low-energy symmetries in
the two systems are at the heart of the different roles of
the SPT anomaly in the two systems.
Though the translation symmetry of the lattice model at the low-energy
limit may look identical to an on-site symmetry of some SPT phase,
the translation symmetry is intrinsically non-on-site.
Hence, it can never be gauged in the precise manner,
Hence, the SPT anomaly, an obstruction of gauging global symmetries,
may not have any implication on the
``non-perturbative'' nature of the theory emergent from the lattice.
(In this context, it may be interesting to ask:
When it is possible to gap the system trivially (i.e., anomaly-free),
is there any way one can adiabatically deform the system to make translation symmetry on-site?)  

%On the other hand, the chiral anomaly does
%not explicitly assume ``gauging''
%of the translation symmetry but assumes only the conserved quantity under
%the translation symmetry. Nevertheless, we show that it
%signals certain amount of the stability of the theory.

%Thirdly, using these understandings of LSMOH theorem in terms of the anomalies, we will extend the con-
%ventional LSMOH theorem, concerning a single particle
%species in general, to the multiple particle species with
%the different charge assignments, e.g., a mixed system of
%charge-2 spinless boson at the half-filling with charge-1
%spinless electron at integral fillings. We construct several
%novel symmetric insulators, which cannot be adiabatically deformed into a Slater-type insulator.

%\subsection{perturbative v.s. non-perturbative stabilities}

%
%We will argue that the above ``subtlety'', which is related to the ``effective'' symmetry,
%is related to the perturbative stability of Fermi surfaces. 
%In contrast to the chiral anomaly, we will argue that the
%SPT anomaly (in the absence of the chiral anomaly)
%implies only the ``local'' stability of the theory emergent
%from the (1+1)d lattice system, i.e., the theory is stable
%to be gapless only ``locally'' near the theory within the parameter space.
%
%
In Sec.\ \ref{(3+1)d Chiral Anomaly and Weyl Semimetals},
we will also consider the (3+1)d chiral anomalies and
relate the anomaly to the ``perturbative" stability.
In contrast to the (1+1)d chiral anomaly detecting
the no-go conditions for the LSMOH theorem,
we show that (3+1)d chiral anomaly only detects the stability of
the gaplessness only near the low-energy theory in the parameter space.

\subsection{Summary}

The above considerations can be summarized, 
from the view point of continuum field theories, 
as follows. 
Let us consider a $(d+1)$-dimensional continuum field theory $\mathcal{F}$.
To be concrete, we assume $\mathcal{F}$ be a theory of {\it relativistic} fermion.
This theory may or may not arise as a low-energy effective theory
of a given lattice model of the same spacetime dimension. 
Let there be a global symmetry $G$
respected by $\mathcal{F}$.
Let there be a 't Hooft anomaly for $G$.
The 't Hooft anomaly has a one-to-one correspondence with
$\Omega^{d+1,\mathrm{tors}}_{\mathrm{Spin}/\mathrm{Spin}^c}(BG)$,
the Pontryagin dual of the torsion subgroup of the equivariant spin/spin$^c$ bordism groups with the symmety group $G$.
\cite{Kapustin2014c}
%$\Omega^{Spin_c}_{d+1}(BG)$.
%$\Omega^{Spin}_{d+1}(BG)$.
Taking this 't Hooft anomaly ``naively'',
one would conclude that the theory must be realized as
a boundary theory of a $(d+2)$-dimensional bulk theory. 

Let us now assume that we actually know that  
$\mathcal{F}$ is a low-energy effective theory
of a $(d+1)$-dimensional lattice model.
Then, at least one of the following must be true:
(a) $G$ is not on-site for the $(d+1)$-dimensional lattice model.
(b) $G$ is not the true symmetry of the problem; It is a symmetry
only emergent in the low-energy physics.
These two possibilities correspond to the LSMOH theorem 
and Fermi surfaces. 
In addition, it should be also noted that, in particular for the case of the LSMOH theorem,
there is no reason to consider relativistic fermions to start with.  
In other words, we need to care about the high-energy scale origins of the
low-energy symmetries
and interpret the meaning of the anomalies carefully.
%At the heart of these discussions, the non-on-site-ness of the translation symmetries play the central roles. 

The rest of the paper is organized as follows.

- In Section \ref{2d SPTs and 1d lattice models}, 
we first review briefly the edge of the 2d quantum spin Hall
effect (QSHE) and show that the exactly same low-energy theory
can arise from the 1d lattice model with spinless fermion at fractional
filling.

-In Section 
\ref{Anomaly and LSMOH theorem}
and 
\ref{SPT anomaly and Local Stability}, 
We discuss the implications of the anomalies. 
With the anomaly-based understandings of the LSMOH theorem,
in section \ref{Anomaly and LSMOH theorem},
we will extend the conventional LSMOH theorem, concerning singly-charged particles in general, to the multiple particle species with the different charge assignments, 
e.g., a mixed system of charge-2 spinless boson at the half-filling with charge-1 spinless electrons at integral fillings. We construct several novel symmetric insulators, which cannot be adiabatically deformed into a Slater-type insulator. 

- In Section \ref{(3+1)d Chiral Anomaly and Weyl Semimetals},
we consider the (3+1)d chiral anomalies\cite{nielsen1983adler,adler1969axial,bell1969pcac} in Weyl and Dirac semimetals\cite{Young2012dirac,yang2014classification,yang2015topological,burkov2011weyl,yang2011quantum,cho2011possible} and relate the anomaly to the ``local" stability. In contrast to the (1+1)d chiral anomaly detecting the no-go conditions for the LSMOH theorem, we show that (3+1)d chiral anomaly only detects the stability of the gapless-ness only near the low-energy theory in the parameter space. We apply these results to the (3+1)d Weyl and Dirac semimetals. 
%We next discuss the role of the (3+1)d chiral anomalies in various topological semimetals in section \textbf{III}.

- We finish by providing conclusions and outlooks in Section \ref{Conclusion and Outlooks}.

\textit{- Note Added}: After the completion of the work,
we became aware of the work by Jian, Bi, and Xu,\cite{jian2017lieb} in which the similar consideration is made.

\section{2d SPTs and 1d lattice models} 
\label{2d SPTs and 1d lattice models} 

In this section, we consider 1d lattice models,
which are enforced to be critical by the
LSMOH theorem, and compare their low-energy theory
to the edge theories of 2d SPTs.
We will find that the lattice model gives rise to exactly the same effective
low-energy theory as the edge of the SPTs.
We will discuss the quantum anomalies relevant for the low-energy theories.  

\subsection{(2+1)d QSHE}
We start by revisiting the simplest (2+1)d fermionic SPT phase, the QSHE,
protected by unitary onsite $U(1)_Q \times U(1)_{S_z}$ symmetry.
The 2d bulk of this SPT phase can be constructed on the honeycomb lattice
by following Kane and Mele\cite{Kane2005}
\begin{align}
  H = -t \sum_{\langle \bm{r}, \bm{r}' \rangle} \Psi^{\dagger}_{\bm{r}}\Psi_{\bm{r}'}
  + i\lambda \sum_{\langle\!\langle \bm{r}, \bm{r}' \rangle\!\rangle}  \Psi^{\dagger}_{\bm{r}} \sigma_z \hat{z} \cdot (\hat{d}^{1}_{\bm{r}\bm{r}'} \times \hat{d}^{2}_{\bm{r}\bm{r}'}) \Psi_{\bm{r}'}, \nonumber
\end{align}
where $\bm{r}$ labels the lattice site and
$\Psi_{\bm{r}} = (c_{\bm{r} \uparrow}, c_{\bm{r}\downarrow})^{\text{T}}$ a spinful fermion;
$\hat{d}^{1,2}_{\bm{r}\bm{r}'}$ is a vector connecting the
next nearest-neighbor sites $\bm{r}$ and $\bm{r}'$ on the honeycomb lattice. 
The lattice Hamiltonian clearly respects the symmetry $U(1)_Q \times U(1)_{S_z}$ at the ultraviolet (UV) scale 
\begin{align}
U(1)_Q:& \quad \Psi_{\bm{r}} \to e^{i \phi} \Psi_{\bm{r}}, \nonumber\\ 
U(1)_{S_z}:& \quad \Psi_{\bm{r}} \to e^{i \sigma_z \theta/2  } \Psi_{\bm{r}}.  
\label{QSHE:Sym}
\end{align}
The ground state is simply the combinations of the completely filled
Chern band with Chern number
$\nu =1$ for spin-$\uparrow$ electrons
and the completely filled Chern band with $\nu =-1$ for spin-$\downarrow$ electrons.

When the open boundary condition
is imposed, 
along a spatial direction $x$, say,
gapless edge states emerge.
They can be described by the low-energy Hamiltonian
\begin{align}
H = \int dx ~ \Psi^{\dagger} (x) (-i v_F \partial_x) \sigma_z \Psi(x),  
\label{QSHE:Edge}
\end{align}
where
$\Psi (x) = (\psi_{\uparrow} (x), \psi_{\downarrow}(x))^{\text{T}}$
and $v_F$ is the Fermi velocity.
The action of the $U(1)_Q \times U(1)_{S_z}$ symmetry on the edge mode
is still given by \eqref{QSHE:Sym},
if the fermionic operators there are replaced by their boundary counterparts. 
It is straightforward to check that there is no gapping term
when the symmetry \eqref{QSHE:Sym} is strictly imposed on the edge. Hence the gaplessness of the edge theory is protected by the symmetry. 

To facilitate
to establish a connection
with filling-enforced gapless states on the 1d lattice,
we note that the criticality
(as well as the quantum anomaly)
of the 1d edge survive even if we lower $U(1)_{S_z}$ down to $\mathbb{Z}_{S_z}$,
i.e., instead of $U(1)_{S_z}$, we can consider
the discrete spin rotation
\begin{align}
  \mathbb{Z}_{S_z}&: \quad \Psi_{\bm{r}} \to e^{i m\theta_F \sigma_z} \Psi_{\bm{r}},
                    \quad  m \in \mathbb{Z},   
\label{QSHE:Sym2}
\end{align}
with some $\theta_F \in (0, 2\pi)$. When we fine-tune $\theta_F =\pi/N$,
we can further ``lower'' the symmetry, $\mathbb{Z}\to \mathbb{Z}_{2N}$. 
%However, by recombining the spin-rotation with $U(1)$, we can restore ``$\mathbb{Z}$-ness" of the spin rotation where the redefined spin rotation comes along with the U(1) global phase rotation 
%\begin{align}
%\mathbb{Z}_{S_z}'&: \Psi_{\bm{r}} \to e^{i \theta_F \sigma_z + i \delta \phi} \Psi_{\bm{r}}. 
%\end{align}
Note also that at the level of non-interacting fermions,   
the classification is still $\mathbb{Z}$ since one can easily verify
that there is no mass term allowed to the theory \eqref{QSHE:Edge}.

\subsection{(1+1)d lattice spinless fermions}

We now consider the model of spinless fermions hopping on
a 1d lattice consisting of $L$ lattice sites:
\begin{align}
  H = -t \sum_{x}^{L} (c^{\dagger}_x c^{\ }_{x+1} + h.c.)
  - \mu \sum_x^{L} c^{\dagger}_x c^{\ }_x.  
\label{Lattice:H}
\end{align}
The model is invariant under the charge $U(1)_Q$,
and, with periodic boundary condition,
lattice translation symmetry $\mathbb{Z}_L$ with $L \gg 1$.
In the thermodynamic limit $L \to \infty$, we have the two symmetries
\begin{align}
U(1)_Q &: \quad c_x \to e^{i\phi} c_x, \nonumber\\ 
\mathbb{Z}_{trans} &:\quad  c_x \to c_{x+1}.  
\label{Lattice:Sym1}
\end{align}
Note that the translation symmetry $\mathbb{Z}_{trans}$ is manifestly non-onsite
at this UV scale.

The ground state can be easily found by filling the single particle states
below the chemical potential $\mu$, 
\begin{align}
 | GS  \rangle \propto \Big(\prod_{|k| \leq k_F} c^{\dagger}_k \Big)|\text{vac} \rangle, \nonumber
\end{align} 
where $|\text{vac}\rangle$ is the Fock vacuum.
The system realizes gapless metal,
which can be easily seen from the band
structure of \eqref{Lattice:H},
if the filling $\nu = \frac{k_F}{\pi} \notin
\mathbb{Z}$.
Furthermore,
the LSMOH theorem\cite{Oshikawa2000} in 1d dictates
that if the translation symmetry $\mathbb{Z}_{trans}$ and $U(1)_Q$ are not broken, then the ground state should be always gapless even in the presence of interactions; It is a filling-enforced critical state. 

To reveal the connection between this 1d lattice model and the edge of
the QSHE, we now proceed to the continuum IR limit of the theory 
\eqref{Lattice:H}
\begin{align}
H = \int dx ~ \Psi^{\dagger} (x) (-i v_F \partial_x) \sigma_z \Psi(x),  
\label{Lattice:LowH}
\end{align}
where $\Psi(x) = (\psi_R (x), \psi_L (x) )^{{T}}$ is the low-energy fermion
field near the Fermi point.
Here, the microscopic fermion operator $c_x$ can be expanded
in terms of the slowly varying low-energy fields $\psi_{R/L}$ as 
\begin{align}
  c_x \approx \psi_R (x) e^{i k_F x} + \psi_L (x) e^{-ik_F x}.
\end{align}
Here we take a convention that the $k_F$ is the right-most momentum of the filled state. 

The symmetry actions of $U(1)_Q \times \mathbb{Z}_{trans}$
within this low-energy theory \eqref{Lattice:LowH} can be easily derived, 
\begin{align}
  U(1)_Q &: \quad \Psi(x) \to e^{i\phi} \Psi(x),
           \nonumber\\ 
\mathbb{Z}_{trans} &: \quad \Psi(x) \to e^{i k_F \sigma^z} \Psi(x).
\label{Lattice:Sym2}
\end{align}
Here, 
in the continuum (conformal) limit
where the UV cutoff, i.e., the lattice constant, is completely ignored,
the translation symmetry $\mathbb{Z}_{trans}$,
which is non-on-site at UV scale,
acts as if it is a purely local, on-site, symmetry
on the infrared (IR) field $\Psi(x)$.
%We can further show that the IR theory \eqref{Lattice:LowH}
%with the symmetry \eqref{Lattice:Sym2}
%suffers from exactly the same anomaly \eqref{Lattice:Anomaly} as its UV
%counterpart, \eqref{Lattice:H},
%i.e., the anomaly is identical at UV and IR scales (see appendix
%\ref{Oshikawa2}).

In summary, 
the continuum IR limit \eqref{Lattice:LowH}
with the symmetry $U(1)_Q \times \mathbb{Z}_{trans}$ \eqref{Lattice:Sym2}
is identical to the edge theory \eqref{QSHE:Edge}
of the QSHE with the symmetry $U(1)_Q \times \mathbb{Z}_{S_z}$
(upto the Fermi velocity
which is irrelevant for the discussion of quantum anomalies).
% Remarkably,
In the IR limit,
the two theories realize the same $\mathbb{Z}$-symmetry actions
(or appropriate subgroup of $\mathbb{Z}$ if the filling is rational fraction
-- see below)
although the symmetry has very different origins at the UV scales.
Furthermore,
on the lattice scale, the translation symmetry is non-on-site (non-local)
\eqref{Lattice:Sym1},
but becomes local \eqref{Lattice:Sym2} on the IR scale
and looks like the onsite internal symmetry
(spin rotation symmetry) of the edge of the 2d QSHE.
In other words, 
the 1d spinless fermion lattice model at fractional filling
evades the no-go theorem
This is exactly parallel
to the proposed dual description of the
half-filled Landau level,
which turns out to be identical to the (dual) low-energy theory of the 2d
boundary of the 3d topological insulator.
%In these descriptions, the symmetries of the low-energy theory is realized
%differently at the UV scales in the two cases,
%though they look identical in the IR limit.
%The topological insulator has local time-reversal symmetry, which is a
%well-defined local symmetry at the UV scale.
%On the other hand, the half-filled Landau level has the particle-hole symmetry,
%which acts locally in the IR limit and identically to the time-reversal symmetry
%of the 3d SPT, and charge conservation.
%However, the particle-hole symmetry cannot be encoded locally at the UV scale in
%the half-filled Landau level.
%Here, inspired by the relation between the half-filled Landau level and the
%surface of topological insulators,
%we have provided a concrete example which circumvents the no-go ``theorem" by
%deriving the low-energy theory
%as well as the symmetry actions at the IR scale from the 1d lattice model and matching them to those of the 1d edge theory of the 2d SPT.

\subsection{Effective symmetry}

It is also important to note that 
when the filling is rational, e.g., $\nu ={1}/{q}$,
then the translation symmetry in Eq.\ \eqref{Lattice:Sym2}
can be reduced to an effective $\mathbb{Z}_{2q}$.
If the center of momentum is shifted to $\frac{\pi}{2q}$ (by some fine-tuning of the band structures), then the translation symmetry is in fact reduced further to $\mathbb{Z}_q$ such that  
\begin{align}
U(1)_Q &:\quad \Psi(x) \to e^{i\phi} \Psi(x), \nonumber\\ 
\mathbb{Z}_q &:\quad  \Psi(x) \to e^{i \frac{\pi}{q} (\sigma^z-1)} \Psi(x). 
\label{Lattice:Sym3}
\end{align}
Thus, the lattice translation symmetry is effectively lowered
to $\mathbb{Z}_q$ -- this reduced symmetry will be called effective symmetry. 

At the free fermion level and within the low-energy theory \eqref{Lattice:LowH},
both the symmetry groups $U(1)_Q \times \mathbb{Z}$ and $U(1)_Q \times
\mathbb{Z}_q$
can protect the gapless-ness of the theory.
At this stage, the lowering of the symmetry as well as treating the translation symmetry as the on-site symmetries are seemingly innocuous. 
However, when it comes to the multiple copies and interactions,
then we will see that these treatments may give rise to subtle effects,
i.e., it now matters if the system comes from the lattice
or from the edge and if the symmetry in the low-energy limit is effective, as we will see from the discussions of the anomaly.

\subsection{Other examples}
Evading the no-go ``theorem'' is also possible in higher dimensions.
For example, the ``duality'' between the composite Fermi liquid in the half-filled Landau
level and the (2+1)d boundary of (3+1)d topological insulators
has been discussed extensively recently.
\cite{Son2015, Geraedts2016, Wang2015CFL,Metlitski2015,Mross2016,Wang2016half}
In this example, the low-energy theory of the half-filled Landau
level is claimed to be the same (in terms of the parity
anomaly, field contents, and symmetry actions) as that
of the surface of 3d topological insulators. Both of these
theories contain a dynamical gauge field and a single
Dirac fermion with the anti-unitary symmetry.
The anti-unitary symmetry looks local in space when acting on the
low-energy fermion fields. This seems against the no-go
theorem since the Landau level can be constructed from
two-dimensional lattices (with the projection to the Landau
level). However, the anti-unitary symmetry
in the Landau level, which emerges after the projection
to the lowest Landau level, is local but non-on-site at lattice scales
(although it acts like a local symmetry in the low-energy field theory description).

It is also instructive to contrast our work with those which deal with weak SPT phases,
e.g., 
Ref.\ \onlinecite{Cheng2015}.
Ref.\ \onlinecite{Cheng2015} considered
$(d-1)$-dimensional lattice models as the surface of $d$-dimensional weak SPT
phases,
% This approach then leads to
and then finds the classification of possible topological orders respecting the translational symmetries from
the $d$-dimensional SPT index,   
$
\mathcal{H}^{d+1}[\mathbb{Z}^{d}_{trans} \times G, U(1) ]
$.
Here $\mathbb{Z}^{d}_{trans}$ is the translation symmetry, and $G$ is the on-site symmetry. 
Through the Kunneth formula, it is found that this index for the weak SPT can be
given in terms of those of the lower-dimensional strong SPTs
with on-site symmetry $G$, i.e.,  
$
\mathcal{H}^{r+1}[G, U(1)] 
$
$(r<d)$, 
which is stacked inside the weak SPT.
Though this formula helps to understand the
index of the weak SPT clearly, this treats the translation symmetry physically
different from the on-site symmetry.
The anomalous nature of the symmetrically-gapped phases of the
$(d-1)$-dimensional lattice models
manifests as the non-trivial indices for this weak SPT phase.
On the other hand, we will consider e.g. the $d=2$ SPT case, or
one-dimensional lattice models, which are forced to be critical instead of
gapped;
The translation symmetry in the lattice models are interpreted as
the on-site symmetry in the strong SPT side.
This makes the non-on-site translation symmetry and on-site global symmetry,
e.g., charge conservation,
of the lattice model to be treated on an equal footing in the SPT side.
Furthermore,
the anomalous nature of the lattice models manifests as the proper
generalizations of the chiral anomaly,
which are a ``more historic" diagnosis of gapless theories than the indices.
Currently, the link between the weak SPT index and the anomaly discussed in this
paper is not obvious,
and thus clarifying the relationships between the cohomological indices $\mathcal{H}^{d+1}$ and chiral anomalies in one-dimensional lattice models and three-dimensional relativistic semimetals will be an interesting future problem.

\section{Anomaly and LSMOH theorem}
\label{Anomaly and LSMOH theorem}

Having confirmed that
the 1d lattice model and the SPT boundary
are described by the same low-energy effective theory
with the identical action of the global symmetry $U(1)_Q\times \mathbb{Z}_N$,
we now proceed to discuss quantum anomalies.  
Here, as emphasized in the previous section,
one should keep in mind the different origins of
the $\mathbb{Z}_N$ symmetry in the LSMOH and SPT contexts.
Nevertheless, in the following we will first
take the low-energy theory on its own
as a relativistic quantum field theory, 
without asking how it arises. 
An obstruction to gauge this global symmetry,
i.e., 't Hooft anomaly
would be labeled by the elements in the three-dimensional equivariant spin$^c$ cobordism group with $\mathbb{Z}_N$ symmetry
\footnote{Here $\Omega^{3}_{\mathrm{Spin}^c}(B\mathbb{Z}_N)=\Omega^{3,\mathrm{tors}}_{\mathrm{Spin}^c}(B\mathbb{Z}_N)$, since the three-dimensional spin$^c$ cobordism group $\Omega^{3}_{\mathrm{Spin}^c}(pt)$ is trivial.} 
\begin{align}
\Omega^3_{\mathrm{Spin}^c}(B\mathbb{Z}_N)\cong\mathbb{Z}_{\epsilon_N\cdot N}\times\mathbb{Z}_{N/\epsilon_N},
\end{align}
where $\epsilon_N=1$ for odd $N$ and $\epsilon_N=2$ for even $N$. (For the known results of $\Omega^3_{\mathrm{Spin}^c}(B\mathbb{Z}_N)$, 
see Ref.\ \onlinecite{Bahri-Gilkey1987, Gilkey1989}.
)
If the low-energy theory is interpreted as the SPT boundary,
$\Omega^3_{\mathrm{Spin}^c}(B\mathbb{Z}_N)$
agrees with the classification of the bulk SPT phases
protected by the unitary on-site symmetry
$U(1)_Q\times \mathbb{Z}_N$.

More concretely, let us consider the following, 
slightly more extended version of
the (1+1)d continuum theory with
the $U(1) \times \mathbb{Z}_N$ global symmetry:
\begin{align}
  H = \int dx\, \sum_{a=1}^{N_f} \Big[
   \psi^{\dagger}_{L,a}i\partial_x\psi_{L,a}
  -
  \psi^{\dagger}_{R,a}i\partial_x\psi_{R,a}
  \Big], 
\label{low-energy_Dirac}
\end{align}
with $N_f$ the number of species.
Henceforth, the velocity is scaled to $1$ for simplicity.
We encode the $U(1)_Q$ symmetry into the fermion fields as
\begin{align}
  U(1)_\epsilon :& \quad
                  \psi_{R,a}(x) \to e^{i \epsilon q_a} \psi_{R,a}(x), \nonumber\\ 
&\quad \psi_{L,a}(x) \to e^{i\epsilon q_a} \psi_{L,a}(x),   
\end{align} 
in which the fermion field $\psi_{R/L, a}$ carries the odd integer electric charge $q_a$. 
In particular, for the electronic systems, all $q_a = 1$. 
On the other hand, the $\mathbb{Z}_N$ symmetry
acts as 
\begin{align}
\mathbb{Z}_{N} :& \quad \psi_{R,a} (x) \to e^{2\pi i s_{R,a}/N} \psi_{R,a} (x) \nonumber\\ 
&\quad \psi_{L,a} (x) \to e^{ 2\pi i s_{L,a}/N} \psi_{L,a} (x), 
\label{low-energy_Dirac_Sym}
\end{align}
with $s_{R/L, a} \in \mathbb{Z}$. 
The full 't Hooft anomalies of $U(1) \times \mathbb{Z}_N$ in the relativistic field theory \eqref{low-energy_Dirac}
can be calculated explicitly,
and are characterized by the following two indices
\begin{align}
  &
  \sum_a\nu_a\cdot\frac{s_{R, a}+s_{L, a}}{\epsilon_N } \mod\mathbb{Z},
  \nonumber \\
  &
\sum_a\nu_a q_a \mod\mathbb{Z}.
    \label{two anomaly indices}
\end{align}
This matches with
$\Omega^3_{\mathrm{Spin}^c}(B\mathbb{Z}_N)\cong\mathbb{Z}_{\epsilon_N\cdot N}\times\mathbb{Z}_{N/\epsilon_N}$.
As we will see in details, the second index is related to
the familiar chiral anomaly in (1+1)d.

As the simpler field theory discussed in the previous section,
the field theory \eqref{low-energy_Dirac}
can be interpreted as either describing a SPT boundary with $U(1)_Q\times \mathbb{Z}_N$
symmetry, or the low-energy effective theory of a 1d LSMOH lattice model.
If the theory originates from the 1d lattice
then the filling $\nu_a$ of the $a$-th fermion
per unit cell fixes the relative difference between the two Fermi points,
i.e., $k_{F,R}^{a} = 2\pi s_{R,a}/N$ and $k_{F,L}^{a} = 2\pi  s_{L,a}/N$, by 
$\nu_a = (k_{F,R}^{a} - k_{F,L}^{a})/2\pi$. The specific positions of $k_{F,R/L}^{a}$ in momentum space depend on the band structure. 

Once we specify the microscopic origin of the field theory
\eqref{low-energy_Dirac},
the two anomaly indices \eqref{two anomaly indices}
have to be interpreted properly. 
As mentioned already, if the low-energy theory \eqref{low-energy_Dirac}
is interpreted as a SPT boundary,
If the low-energy theory is interpreted as the SPT boundary,
$\Omega^3_{\mathrm{Spin}^c}(B\mathbb{Z}_N)$
agrees with the classification of the bulk SPT phases
protected by the unitary on-site symmetry
$U(1)_Q\times \mathbb{Z}_N$.
%{\color{red}
%  The first index is the SPT anomaly}
%and the second index is the chiral anomaly.
If the symmetries were realized on the SPT boundary,
then any of non-zero anomalies implies a non-pertubative obstruction
for a symmetric trivial gapped state at the boundary.

On the other hand, as we will argue,
in the LSMOH context, only the second index, the chiral anomaly,
is relevant;
The second index is (a slight extension of)
the conventional LSMOH theorem.
In the usual LSMOH theorem, the charge is taken to be $1$, i.e., all $q_a =1$. Then the absence of the chiral anomaly with $q_a =1$ for all $a$ is equivalent to the conventional LSM theorem. However, here we allow generic $q_a$ here (we assume the existence of the minimal charge-$1$ fermion in the spectrum). We will include the bosons later in the section \ref{LSM} by using the momentum pumping argument. 

Given the identification of the chiral anomaly to the LSMOH theorem,
the first index must be something intrinsic to the SPT boundaries
where $\mathbb{Z}_N$ is truly on-site, and it is unrelated to the LSMOH theorem.
We call this the anomaly $\mathbb{Z}_N$ anomaly below to distinguish it from the chiral anomaly.
%\textcolor{red}{Hence, we call it as the ``SPT anomaly".}
We will find that it does not gaurantee the non-perturbative stability for the lattice models but may imply only the local stability.

%Below, we explain the roles of each anomaly separately. 

\subsection{Chiral anomaly and the LSMOH theorem}\label{LSM}
We now give some details for the identification of the second index
as the chiral anomaly.
We will also see that it is nothing but the LSMOH theorem.
%Next, we prove that the absence of the chiral anomaly is equivalent to the existence of the symmetric insulator. 

\subsubsection{Chiral Anomaly: Field Theory}
To see the second index is equiavelent to the chiral anomaly
% in the free fermion case.
we ``promote'' the $\mathbb{Z}$ symmetry (or $\mathbb{Z}_N$) to the continuous axial $U(1)_A$ 
and use the standard anomaly equation. 
%first give a simpler physical picture of the chiral
%nomaly and then provide a formal derivation. 
For the case of the single-flavor model, 
the violation of the axial charge conservation
in the presence of the electromagnetic (vector) gauge field
is quantified by
\begin{align}
\frac{dQ_5}{dt} = \partial_\mu j^{\mu}_5 = \frac{1}{\pi} \int dx~ q E_x, 
\end{align}
where 
the axial charge, the number difference between the left mover and the right
mover, corresponds to the momentum because the momentum is $P = k_F
(R^{\dagger}R - L^{\dagger}L) = \nu \pi Q_5$,
where $Q_5$ is the axial charge;
$E_x$ is the electric field along the $x$-direction and $q$ is the
electric charge.
%The chiral anomaly signals the violation of the axial charge conservation
%in the presence of the electromagnetic (vector) gauge field.
%In the low-energy limit, the axial charge, the number difference between the left mover and the right mover, corresponds to the momentum because the momentum is $P = k_F (R^{\dagger}R - L^{\dagger}L) = \nu \pi Q_5$, where $Q_5$ is the axial charge. Now the change in the axial charge satisfies the following 
%\begin{align}
%\frac{dQ_5}{dt} = \partial_\mu j^{\mu}_5 = \frac{1}{\pi} \int dx~ q E_x, 
%\end{align}
%where $E_x$ is the electric field along the $x$-direction and $q$ is the
%electric charge.
When we have multiple species of the fermions $\{ \Psi_a \}$ with multiple charge $q_a$, the total momentum transfer in the system is the sum of the momentum transfer of each species $a$, i.e., 
\begin{align}
\frac{dP_{tot}}{dt} = \sum_a \nu_a \pi \frac{dQ_{5,a}}{dt} = \sum_a \nu_a q_a \int dx~ E_x. 
\end{align}
Now we imagine to thread a magnetic flux $\int dx~A_x = 2\pi$ adiabatically to the system. Then during the process, the change in the momentum is 
\begin{align}
\triangle P_{tot} = \sum_a \nu_a q_a \int dx dt ~ \partial_t A_x = 2\pi \sum_a \nu_a q_a. 
\end{align} 
Hence, the momentum (axial charge) is not conserved
during the process if $\triangle P_{tot}$ is non-zero.

Remembering now that the relevant symmetry is 
 $\mathbb{Z}$ (instead of the axial $U(1)_A$), 
and that $\mathbb{Z}$ is derived from
lattice translation symmetry,
we know that $P_{tot}$ is conserved only modulo $2\pi$.
%Then, $Q_5$ is not conserved when $\triangle P_{tot}$
%is not a multiple of $2\pi$.
Hence, the anomaly-free condition is given by 
\begin{align}
\sum_a \nu_a q_a = 0 \quad \mod \mathbb{Z}. 
  \label{extended LSM}
\end{align}
Otherwise, the translation symmetry and the charge conservation are in conflict.

The above result can also be derived in a more general and formal setting. 
%With this physical picture in hand, we now present a formal derivation.
Consider (intrinsically-continuum) Dirac fermions
which carry odd electric charges
in the presence of the background $U(1)$ gauge field on a closed manifold $M$.
Specifically, we formulate the fermion theory \eqref{low-energy_Dirac} on a
generic closed Riemannian two-manifold $(M, g)$ endowed with a $\mathrm{spin}^c$
structure, where a well-defined $\mathrm{spin}^c$ connection, denoted as $A$,
exists. 
(Only for this part of the discussion,
we temporarily restrict fermions to carry
odd charge $q_a \in 2\mathbb{Z} +1$
(and bosons to carry even charge) to use
the $\mathrm{spin}^c$ connection.)
Here we work in Euclidean signature.
[At this stage, one may worry that we are imposing two much 
structures (e.g., spin structure and continuum manifold with metric $g$), unrelated to the lattice fermions. However, these are only for the concreteness.]
Since there is no (gauge) anomaly for ${U}(1)_Q$
%(as the low-energy theory is described in terms of Dirac fermions),
the partition function $Z_{\{\Psi_a\}} (M; g, A)$ is well-defined on any such
two-dimensional $\mathrm{spin}^c$ manifold $M$. However, $Z_{\{\Psi_a\}} (M; g,
A)$ might in general not be invariant under $G_{eff}= \mathbb{Z}_q$ (or $G=
\mathbb{Z}$),
which is a symmetry of the classical action of \eqref{low-energy_Dirac}
with background gauge field $A$.
This is a discrete analog
% -- as $G_{eff}$ is discrete --
of the usual chiral (axial) anomaly of the continuous axial symmetry for Dirac fermions, and one can similarly use Fujikawa's method to compute such a discrete chiral anomaly 
The anomaly comes from the nontrivial transformation of the path integral measure 
$\prod_a\mathcal{D}\Psi_a\mathcal{D}\bar{\Psi}_a$ under $G_{eff}$: 
\cite{araki2008non,ishimori2010non}
\begin{align}
&Z_{\{\Psi_a\}} (M; g, A) 
\nonumber\\
&\overset{G_{eff}}{\longrightarrow}
 \exp\left(2\pi i\sum_a\nu_a \mathcal{I}_{q_a}(M; g, A)\right) Z_{\{\Psi_a\}} (M; g, A),
\end{align}
where $\mathcal{I}_{q_a}(M; g, A)$ is the index of the charge-$q_a$ Dirac operator $i\slashed{\partial}-q_a\slashed{A}$ on $M$ endowed with a metric and a $\mathrm{spin}^c$ structure.
Since any two-dimensional $\mathrm{spin}^c$ manifold $M$ is bordant to a multiple of $\mathbb{CP}^1\cong S^2$,
\footnote{
To be more precise, any two-dimensional $\mathrm{spin}^c$ manifold $M$, together with its given $\mathrm{spin}^c$ structure, is bordant to a multiple of $\mathbb{CP}^1$ with a (canonical) $\mathrm{spin}^c$ structure.
}
that is, $[M] = k_M\cdot [\mathbb{CP}^1]$ for some integer $k_M$,
where $[\ \cdot\ ]$ denotes the equivalence class under bordism, and the index  $\mathcal{I}_{q_a}(\ \cdot\ )$ is a bordism invariant, we have
\begin{align}
  \mathcal{I}_{q_a}(M; g, A)
  &= k_M \cdot \mathcal{I}_{q_a}(\mathbb{CP}^1; g', A')
    \nonumber \\
 &=  k_M \cdot q_a\int_{S^2} c_1 (F) 
\nonumber\\
&= - k_M q_a,
\end{align}
where $c_1(F)$ is the first Chern class of the field strength $F=dA$ of a $\mathrm{spin}^c$ connection $A$ on $\mathbb{CP}^1$.
Then, it is obvious that the anomaly-free condition for the theory on any $M$ is given by
\begin{align}
\exp\big(-2\pi ik_M\sum_a\nu_a q_a\big)=1 \ \
\Longleftrightarrow
\ \ \sum_a\nu_a q_a \in\mathbb{Z}.
\end{align}

%Given the derivations, one may ask if the effect of the interactions on this
%chiral anomaly.
%In principle, the anomaly is expected to be preserved as far as the spectrum
%remains gapless.
%In fact, this can be explicitly shown by using the bosonized representation
%of the fermions where we deal with the (strong forward-scattering) interactions
%non-perturbatively and so we do not need to assume the free fermion systems.
%The result remain the same and we put this discussion into the appendix \ref{chiral}.  

The chiral anomaly or the momentum pumping can be also computed by using bosonization. 
Let us again start from the low-energy Hamiltonian
\begin{align}
H = \int dx ~ \Psi^{\dagger} (x) (-i \partial_x) \sigma_z \Psi(x),  
\end{align}
where $\Psi (x) = (R(x), L(x))^{\text{T}}$.
We first need to implement the twisted boundary condition by
$U(1)_Q$ on the circular edge $x \sim x+L$ as
\begin{align}
\Psi(x) = e^{-i\phi_Q} \Psi(x+L).
\label{QSHE:BC}
\end{align}
We call the resulting ground state
in the presence of this boundary condition as $|\phi_Q \rangle$. 

A convenient way to construct and study $|\phi_Q\rangle$ 
is to (abelian) bosonization.
By bosonization, we represent
the fermionic operators as $R \sim e^{i\phi_\uparrow}$ and $L \sim e^{i\phi_\downarrow}$ with the following commutators 
\begin{align}
[\phi_{\sigma'}(x'), \partial_x \phi_{\sigma}(x)] = 2\pi i \cdot \text{sgn}(\sigma) \delta_{\sigma,\sigma'} \delta(x-x'), 
\end{align}
where $\text{sgn}(\uparrow) = +1$ and $\text{sgn}(\downarrow) = -1$.
Correspondingly, the densities of $\psi_\uparrow$ and $\psi_\downarrow$ are
given by $\rho_\uparrow(x) = \frac{1}{2\pi} \partial_x \phi_\uparrow (x)$ and
$\rho_\downarrow (x) = -\frac{1}{2\pi} \partial_x \phi_\downarrow (x)$.
The conserved charge and the momentum 
can be constructed as
$Q = Q_\uparrow + Q_\downarrow$,
$P = \nu \pi (Q_\uparrow - Q_\downarrow)$,  
where $Q_\uparrow = \int \rho_\uparrow (x) dx$ and $Q_\downarrow = \int \rho_\downarrow (x) dx$. 

The ground state $|\phi_Q\rangle$
in the presence of the twisted boundary condition
$
\Psi (x) = e^{-i\phi_Q} \Psi (x+L)
$ 
obeys 
\begin{align}
\Big( \psi_\sigma (x) - e^{-i\phi_Q} \psi_\sigma (x+L) \Big) |\phi_Q \rangle = 0. 
\end{align} 
By the standard operator-state correspondence in CFT,\cite{PolchinskiStringbook,Ginsparg91,CFTbook} we can represent such state by 
\begin{align}
&|\phi_Q \rangle =\lim_{\tau \to -\infty} V_{\phi_Q}(\tau) |0 \rangle,    
\nonumber \\
&V_{\phi_Q}(\tau) \sim e^{i\frac{\phi_Q}{2\pi} (\phi_\uparrow (\tau) - \phi_\downarrow(\tau))},
\end{align}
where $|0\rangle$ is the ground state of the untwisted sector.
Now the (relative) quantum number carried by $|\phi_Q \rangle$ can be directly read off from 
the operator $V_{\phi_Q}$ because
$
[Q, V_{\phi_Q} ]|0 \rangle =  Q V_{\phi_Q}|0\rangle = Q |\phi_Q \rangle,  
$
where we have used $Q |0\rangle = 0$. On the other hand, $[Q, V_{\phi_Q}] = 0$
from the direct computation of the commutator.
Hence, $|\phi_Q\rangle$ does not carry any charge.
On the other hand, one verifies $[P, V_{\phi_Q}] =
\phi_Q \nu $ and $|\phi_Q \rangle$ carries the momentum
$P= \phi_Q \nu $.
Hence, when $\phi_Q = 2\pi$,
the ground state has momentum $2\pi \nu$ relative to the untwisted
sector $\phi_Q = 0$. This is consistent with the previous field theory
calculations.

\subsubsection{Chiral Anomaly: Lattice formulation}
Note that the above discussions rely on the specific assumptions on the ground state: Fermi liquid (or Luttinger liquid). Here, we derive the same anomaly-free condition without assuming a particular ground state. This allows us to include the interacting bosonic case to the discussion, where we do not assume a Fermi-liquid like state. This is a reformulation of Oshikawa's argument \cite{Oshikawa2000}. 

We start from a system of fermions
defined in terms of the fermion annihilation operators
$c_x$ where $x$ is the lattice site.
To follow the field theoretic discussion,
we consider the gauge field $A_x$.
This can be implemented by the boundary conditions by $\hat{T}_1^{L} = e^{i\Phi}
\hat{I}$,
where $L$ is the length of the space and $\Phi$ is the flux threaded into the
space (here $\hat{T}_1$ is the translation symmetry operation acting on the
fermion $c_x$).
Here $\hat{I}: c_x \to c_x$ is the identity operation on the fermion operator
$c_x$.
Then, solving back to $\hat{T}_1$ for this,
we find the (simplest) solution $\hat{T}_{1} (\Phi) : c_x \to c_{x+1} e^{i\Phi/L}$.

Next we assume a translation symmetric ground state at the zero flux sector 
\begin{align}
|GS \rangle = \sum_{j = 1}^{N} A(\{x_j\}) c^{\dagger}_{x_1} c^{\dagger}_{x_2} \cdots c^{\dagger}_{x_N} |0 \rangle,
\end{align}
in which $|0\rangle$ is the Fock vacuum defined in the microscopic Hilbert
space. 
$A(\{ x_j \})$ is the coefficient for the configuration $\{ x_j \}$. 
Now, we act with the translation symmetry at $\Phi = 0$, i.e., 
\begin{align}
  &
  \hat{T}_1 (\Phi = 0) |GS \rangle
    \nonumber \\
  & =
    \sum_{j = 1}^{N} A(\{x_j\}) c^{\dagger}_{x_1+1} c^{\dagger}_{x_2+1} \cdots c^{\dagger}_{x_N+1} |0 \rangle
    \nonumber\\
  &= \sum_{j = 1}^{N} A(\{x_j-1\}) c^{\dagger}_{x_1} c^{\dagger}_{x_2} \cdots c^{\dagger}_{x_N} |0 \rangle.
\end{align}
Because of the translation symmetry, we find that 
\begin{align}
\sum_{j = 1}^{N} A(\{x_j-1\}) c^{\dagger}_{x_1} c^{\dagger}_{x_2} \cdots c^{\dagger}_{x_N} |0 \rangle \nonumber\\  
  = e^{iP_0}\sum_{j = 1}^{N} A(\{x_j \}) c^{\dagger}_{x_1} c^{\dagger}_{x_2} \cdots c^{\dagger}_{x_N} |0 \rangle,
\end{align}
in which $P_0$ is the momentum of the ground state. Now applying the translation symmetry with $\Phi= 2\pi$ flux into the ground state with the assumption that the ground state comes back to itself after insertion of $2\pi$ flux, we find 
\begin{align}
\hat{T}_1 (\Phi = 2\pi) |GS \rangle  = e^{2\pi i \frac{N}{L}} e^{iP_0}  |GS \rangle  = e^{2\pi i \nu} e^{iP_0}|GS \rangle. \nonumber
\end{align}
(Here, $\hat{T}_1 (2\pi)$ is in fact related to $\hat{T_1}(0)$ by a large gauge transformation $U = \exp [2\pi i \sum_x x \hat{n}_x/L]$ such that $\hat{T}_1 (2\pi) = U^{\dagger} \hat{T}_{1}(0) U$, see Ref.\ \onlinecite{Oshikawa2000}.)
Hence the momentum pumped into the system is $2\pi \nu$ when the flux is $2\pi$.
This nicely matches the chiral anomaly calculation.
Note that when we have multiple species of particles,
the phase factors will sum up to each other. Hence, the triviality of the momentum pumping is $\sum_a q_a \nu_a =0$ mod $\mathbb{Z}$. 

It should be also noted that, within this pumping argument, we do not make
an assumption about the statistics of the particles.
Hence, the criteria is now independent of the statistics.
Note also that the minimum transparent flux $2\pi$ is imposed
by assuming the presence of the charge-1 particle in the spectrum, which may be gapped or gapless.

\subsubsection{Extension of the LSMOH theorem}\label{LSMOH_theorem}

From the comparison between the chiral anomaly calculation and the LSMOH
theorem,
we expect that when the anomaly free condition \eqref{extended LSM}
is satisfied, it should be possible to gap the system trivially without symmetry breaking.
%In this section, we show/prove that the absence of the chiral anomaly is the existence of trivial symmetric insulator.
This section is devoted to construct the symmetric insulators explicitly by using bosonization. 
%With the above arguments, we now present our extended LSMOH theorem in 1d with
%the charge conservation and translation symmetry.
%The symmetric insulator is available only when 
%\begin{align}
%\sum_a \nu_a q_a = 0~\text{ mod }~\mathbb{Z}. 
%\end{align}
More specifically, 
let us write $[\nu, q]$ to represent the "equivalence class" of the physical
Hamiltonian of the charge-$q$ particle, either fermionic or bosonic, system at the filling $\nu$, under the consideration based on the (generalized) LSMOH theorem.
Then, we prove the followings: 
\begin{subequations}
\begin{align}
\label{LSM_fermion_chain_a}
[\nu, q]&=0, \quad\text{if}\ \nu q\in\mathbb{Z}; \\ 
\label{LSM_fermion_chain_b}
[\nu_1+\nu_2, q] &= [\nu_1, q] \oplus [\nu_2, q]; \\
\label{LSM_fermion_chain_c}
[\nu, q_1 + q_2] &= [\nu, q_1]  \oplus [\nu, q_2].
\end{align}
\end{subequations}
Here we use $\oplus$ to denote the direct sum of various systems.
We also need to define the trivial class (phase), denoted as $0$ above, as follows:
\begin{itemize}
  \item[(1)] A system that can be gapped (with the consideration of interactions) in a symmetry-preserving fashion is trivial;
\item[(2)]
(``Stably-trivial'' condition) If a system can be gapped,
when coupled to some trivial systems of the first kind, in a symmetry-preserving fashion, then it is also trivial.
\end{itemize}
The properties (\ref{LSM_fermion_chain_a})-(\ref{LSM_fermion_chain_c}) are naturally satisfied from the point of view of anomalies, as $[\nu, q]$ can actually be characterized by the chiral anomaly index $\nu q\mod\mathbb{Z}$. (In the case, "$\oplus$" represents an usual addition of numbers in $\mathbb{R}/\mathbb{Z}$.)
Nevertheless, here we perform a stability analysis to provide another evidence to conform the anomaly argument presented before.

It should be noted that 
the meaning of being trivial in the current context
is different from the SPT context.
First, 
for the non-perturbative stability of lattice systems,
the trivial system is solely identified through their chiral anomaly.
The other part of the 't Hooft anomaly is not important.
In other words, trivial systems in this context may
accidentally have non-trivial $\mathbb{Z}_N$ anomaly
(the other part of the full 't Hooft anomaly).
Second, here we use the full translation symmetry $\mathbb{Z}$,
instead of the effective $\mathbb{Z}_N$ with fixed $N$.
Hence, trivial systems in the present context are
different from trivial systems
in the SPT context where we classify SPT boundaries
with $U(1) \times \mathbb{Z}_N$ with fixed $N$.
The different meaning of trivial states is also reflected
in how we ``add'' (and ``subtract'') systems. 
In the SPT context, we are allowed to add only systems with vanishing
full 't Hooft anomalies,
which should be contrasted with our Condition (2).
Also in the SPT context we do not add trivial systems
with the ``extended'' symmetries than the symmetry of edge theories.
%This makes the difference between the LSM-related
%trivial conditions and the SPT-related trivial conditions. 
In other words,
the label $N$ should not be treated as a fixed symmetry label,
but as indicating a representation under the microscopic
translation symmetry (rather than effective symmetry).

Our result for symmetric insulators,  
$
\sum_a \nu_a q_a = 0
$
$\mod \mathbb{Z}$,  
and the following field theory discussion 
have some resemblance to the recently-discovered lattice homotopy
argument.\cite{po2017lattice}
The statement in the lattice homotopy argument is that,
as far as the lattice symmetry in concern is not changed (in our case, it is the
translation symmetry),
the system can be deformed into a simpler lattice by adding the symmetry charges from each lattice sites. 
For example, imagine a system of two spin-$\frac{1}{2}$'s per unit cell, 
e.g., two lattice sites in the unit cell, 
and we want to impose the translation symmetry only. 
Then, as far as the translation symmetry is concerned, 
we can deform the lattice so that the two lattice sites are sitting on top of each other, 
and combine the two spin-$\frac{1}{2}$ into a single spin-$1$ or spin-$0$ object. 
Then we know that the integral spin inside the unit cell gives a trivial insulator.
In our case, we are adding up the electric charges of particles per unit cell. 
Then the criteria we obtained is equivalent to having an integral charge per unit cell. 
Hence, our anomaly seems to be the manifestation of the lattice homotopy for the translation symmetry case.

%To show
%\eqref{LSM_fermion_chain_a}, \eqref{LSM_fermion_chain_b} and
%\eqref{LSM_fermion_chain_c},
%we go through a few steps. 
%We will use these conditions extensively to prove 
%Eqs.\ \eqref{LSM_fermion_chain_a}, \eqref{LSM_fermion_chain_b}, 
%and \eqref{LSM_fermion_chain_c}.  
%In addition, 
%\eqref{LSM_fermion_chain_a}, \eqref{LSM_fermion_chain_b} and \eqref{LSM_fermion_chain_c}
%are derived with the assumption that there is a charge-1 particle in the spectrum.

\paragraph{Purely Fermionic Case:} 

To show
\eqref{LSM_fermion_chain_a}, \eqref{LSM_fermion_chain_b} and
\eqref{LSM_fermion_chain_c},
we go through a few steps. 
In addition, 
we assume that there is a charge-1 particle in the spectrum.
We first warm up with the fermionic case.
For this case, we use the subscript $F$ to represent the fermion,
i.e., $[\nu, q]_F$. 

\textit{Step 0.}
For the property \eqref{LSM_fermion_chain_a},
it is easy to show that $[\nu, q]_F=0$ when $\nu\in\mathbb{Z}$,
as one can introduce a conventional backscattering term
to have a symmetric gapped ground state.
When $\nu=k/q\notin\mathbb{Z}$,
let us consider the low-energy theory  
\begin{align}
H =  \int dx\,\Psi^{\dagger}_1 (-i\partial_x) \sigma^z \Psi_1.  
\end{align}
Here the $\Psi_1$ is the charge-$q$ fermion field at the filling $k/q$
with the translation symmetry 
\begin{align}
trans:& \quad \Psi_1 \to e^{i\pi k/q \sigma^z + i \bar{k}} \Psi_1. 
\end{align}
Without losing generality, we can take the center of the momentum $\bar{k} = 0$ for this fermionic system. 

To gap out the system,
we couple the system to other two systems at integral filling
which consist of particles carrying unit charge;
both of these systems are trivial.
We then find possible gapping potentials.
For example, we consider 
\begin{align}
[k/q, q]_F\oplus [0,1]_F \oplus [0,1]_F, 
\end{align}
for which the corresponding Hamiltonian is given by 
\begin{align}
H = \int dx\, \sum_{a=1}^{3} \Psi^{\dagger}_a (-i\partial_x) \sigma^z \Psi_a,
\end{align}
where $\Psi_{2,3}$ are the fermion fields from the $[0,1]_F$ sectors.
Under the translations, $trans: \Psi_{2,3} \to \Psi_{2,3}$
because they are at zero filling.
Next we use the bosonization representation
to write $\Psi_1 \sim (e^{i\phi_1}, e^{i\phi_2})$,
$\Psi_2 \sim (e^{i\phi_3}, e^{i\phi_4})$ and so on,
where $\phi_i$ are properly compactified bosonic fields. 
We can then construct bosonic fields
$\Phi_j \sim \exp(i \vec{l}_j \cdot \vec{\phi})$
with $\vec{\phi} = (\phi_1, \phi_2, \cdots, \phi_6)$ 
and 
\begin{align}
\vec{l}_1 &= (1, 1, -q, 0, 0, -q), \nonumber\\
\vec{l}_2 &= (q, -q, 2, -2, 0, 0), \nonumber\\
\vec{l}_3 &= (0, 0, 1, -1, -1, 1).
\end{align}
By adding the interaction term $\propto - \sum_j (\mu_j \Phi_j + h.c.)$,
with large enough $\mu_j$,
which will condense the bosons $\Phi_j$, 
we obtain a gapped ground state which is symmetric.
We have thus shown $[k/q,q] =
[k/q, q]\oplus [0,q] \oplus [0,q] = 0$.
This confirms Eq.\ \eqref{LSM_fermion_chain_a}.

\textit{Step 1.}
Independence to the center of momentum:
For the gappability conditions, we can show that the center of momentum
(i.e., the center between the left-mover's and right-mover's momenta)
is not important. 
Although the center of momentum is $0$ when the lattice
model has accidental parity symmetry,
it needs not to be so.
Here, we will show that the center of momentum can be forgotten
for those conditions and can be taken to be $0$ without losing generality. 

To show this, we need to show that the charge-$q$ fermionic system
$[\nu, q, \bar{k}]_F$
(with the third index to represent the center of the momentum)
at filling $\nu$ with the center of momentum at $\bar{k} \in (-\pi, \pi)$
is equivalent to another charge-$q$
fermionic system $[\nu, q, 0]_F$ at filling $\nu$
with vanishing center of momentum. For this, we show that 
\begin{align}
[\nu, q, \bar{k}]_F = [\nu, q, 0]_F.    
\end{align}
This can be shown by constructing a symmetric gap
for the coupled two fermionic systems,
$[\nu, q, 0]_F$ and $[1-\nu, q, \bar{k}]_F$ with arbitrary $\bar{k}$. 

%Then, this automatically proves that a charge-q fermionic system at the filling $\nu$ with any center of the momentum $\bar{k}$ can be gapped by another charge-q fermionic system of the filling $1-\nu$ whose center of momentum is at $0$ (note that the momentum is relative and can be shifted away by constant always).  
%Now we construct a symmetric gap for the two fermionic systems, $[\nu, q, 0]_F$ and $[1-\nu, q, \bar{k}]_F$ for any $\bar{k}$. For this system, we have 
For the coupled system,
$[\nu, q, 0]_F$ and $[1-\nu, q, \bar{k}]_F$ with arbitrary $\bar{k}$,
let us consider the Hamiltonian 
\begin{align}
H = \int dx\, \sum_{a} \Psi^{\dagger}_a (-i\partial_x) \sigma^z \Psi_a  
\end{align}
with $U(1)_{\delta \phi}: \Psi_a \to e^{i q \delta \phi} \Psi_a$.
The translation symmetry is encoded as 
\begin{align}
  trans:& \quad \Psi_1 \to e^{i\pi \nu \sigma^z} \Psi_1,
          \nonumber\\ 
&\quad \Psi_2 \to e^{i\pi (1-\nu) \sigma^z + i \bar{k}} \Psi_2.   
\end{align}

To construct the symmetric gap,
we use the stably-trivial condition and
include the following two trivial charge-$q$ fermion fields
$\Psi_3$ and $\Psi_4$ which transform under translation as  
\begin{align}
  trans:& \quad \Psi_3 \to e^{-i\pi \sigma^z} \Psi_3,
          \nonumber\\ 
&\quad \Psi_4 \to e^{i\bar{k}}\Psi_4. 
\end{align}
Note that $\Psi_3$ is at filling $\nu =1$ and $\Psi_4$ at $\nu=0$ and thus they are trivial. %Indeed, the conventional backscattering terms $\Psi^{\dagger}_a \sigma^x \Psi_a$ for $a=3,4$ can gap out the two fermionic systems (for $a=3$ and $a=4$) without breaking any symmetry. This proves that the two systems are trivial. 
To gap out $a=1,2,3,4$ all together in a symmetric fashion,
we need to consider the gapping potentials generated by the following vectors 
\begin{align}
&\vec{l}_{1} = (1,0,1,0,0,-1,0,-1), \nonumber\\ 
&\vec{l}_2 = (0,1,0,1,-1,0,-1,0), \nonumber\\ 
&\vec{l}_3 = (1,1,0,0,-1,-1,0,0), \nonumber\\
&\vec{l}_4 = (0,0,1,1,0.0,-1,-1).
\end{align}
The bosons to condense are given
by $\Phi_j \sim \exp(i \vec{l}_j\cdot \vec{\phi})$,
where $\vec{\phi} = (\phi_1, \phi_2, \cdots, \phi_8)$
with the bosonization representation $\Psi_1 \sim (e^{i\phi_1}, e^{i\phi_2})$, $\Psi_2 \sim (e^{i\phi_3}, e^{i\phi_4})$ and so on. We can show that the condensed bosons do not break any symmetry. 

Hence, with this, for the given filling $\nu$ and charge $q$,
we can now take the center of momentum to be zero,
i.e., $\bar{k}=0$, to study the symmetric gappability.
So, from here and on, we drop the index $\bar{k}$
from $[\nu, q, \bar{k}]_F$ to represent systems
and simply write $[\nu, q]_F$.  

\textit{Step 2.}
To verify the property \eqref{LSM_fermion_chain_b},
we show that $[\nu_1, q]_F \oplus[\nu_2, q]_F \oplus[\nu_1+\nu_2,q]^{-1}_F$ is trivial, where $[\nu,q]^{-1}_F=[-\nu,q]_F$ is the inverse of a phase $[\nu,q]_F$
(as $[\nu,q]_F\oplus[-\nu,q]_F$ can be trivially gapped). A simple way to do this is by coupling it to a trivial phase, say, $[0, q]_F$; that is, we instead consider 
\begin{align}
[\nu_1, q]_F\oplus[\nu_2, q]_F\oplus[\nu_1+\nu_2,q]^{-1}_F\oplus[0, q]_F
\end{align}
and examine its stability.
In fact, 
a set of null vectors
(in the above order of the bosonized fields)
can be chosen as
\begin{align}
\vec{l}_1 &= (1, 0, 1, 0, 0, -1, 0, -1), \nonumber\\
\vec{l}_2 &= (0, 1, 0, 1, -1, 0, -1, 0), \nonumber\\
\vec{l}_3 &= (1, 1 ,-1, -1, 0, 0, 0, 0),  \nonumber\\
\vec{l}_4 &= (0, 0, 0, 0, 1, 1, -1, -1),
\end{align}
such that the ground state is symmetry invariant. This confirms Eq. \eqref{LSM_fermion_chain_b}.

\textit{Step 3.}
Finally, the property \eqref{LSM_fermion_chain_c} can be derived in a
similar way.
We consider the following combination: 
\begin{align}
&[\nu, q_1]_F\oplus[\nu, q_2]_F\oplus[\nu, q_1+q_2]^{-1} \oplus [-\nu,0]_F,
\end{align}
where the extra term $[-\nu,0]$ is a trivial phase, as it ca be trivially gapped.
Then a set of null vectors (in the above order) for gapping such a system can be chosen as
\begin{align}
\vec{l}_1 &= (1, 0, 1, 0, 0, -1, 0, -1), \nonumber\\
\vec{l}_2 &= (0, 1, 0, 1, -1, 0, -1, 0), \nonumber\\
\vec{l}_3 &= (1, -1 ,-1, 1, 0, 0, 0, 0),  \nonumber\\
\vec{l}_4 &= (0, 0, 0, 0, 1, -1, -1, 1),
\end{align}
This confirms (\ref{LSM_fermion_chain_c}).

\paragraph{Boson-Fermion Conversion:} 
Next we prove $[\nu, q]_B = [\nu, q]_F$ where the subscript $B$ represents that the system is made of 
bosons. 
To show this, we show 
\begin{align}
[\nu, q]_B \oplus [\nu, q]_{F}^{-1} = [\nu, q]_B \oplus [1-\nu, q]_F =0.  
\end{align}
The low-energy theory of the bosons $[\nu, q]_B$ 
is given by 
\begin{align}
\Phi(x) &\sim e^{i\phi}, 
\nonumber \\ 
\rho(x) &\sim \frac{1}{\pi}\partial_x \theta + \sum_{n\neq 0}\rho_n e^{in(2k_F x + 2\theta) },  
\end{align}
with $2k_F = 2\pi \nu$, and $[\partial_x \theta(x), \phi (x')] = i \pi \delta(x-x')$ 
($\theta$ is compactified as $\theta\equiv \theta+\pi$, i.e., $e^{2i\theta}$ is the smallest local field involving $\theta$). 
Here $\Phi(x)$ represents 
the fundamental local boson field. 

Without losing generality, we can take the low-energy theory of $[1-\nu, q]_F$ whose center of momentum is zero. 
\begin{align}
trans:& \quad \phi_1 \to \phi_1 + \pi (1-\nu),
\nonumber\\ 
&\quad \phi_2 \to \phi_2 - \pi (1-\nu), 
\end{align}
where the fermions are written as $\Psi = (e^{i\phi_1}, e^{\phi_2})$ with the usual kinetic term. Now the symmetric gapping potentials are given by the following two operators $B_i \sim \exp(i \vec{l}_i \cdot \vec{\phi})$ such that  
\begin{align}
&\vec{\phi} = (\phi, \theta, \phi_1, \phi_2), \nonumber\\ 
&\vec{l}_1 = (-2,0,1,1),
\quad 
\vec{l}_2 = (0,2,1,-1).
\end{align}
The vectors $\vec{l}_i$ are chosen so that 
$
\vec{l}_i \cdot K^{-1} \cdot \vec{l}_j = 0
$
with $K^{-1} = (\frac{1}{2}\sigma_x) \oplus \sigma_z$. 

Hence, with this conversion 
between fermions and bosons, 
we can now ignore 
the distinction between fermions and bosons 
in terms of the symmetric gappability. 
Thus, when the filling $\nu$ and the charge $q$ are given, we can in general take the charge-$q$ fermionic system to represent $[\nu, q]$ to investigate the gappability conditions. 
Hence, from here and on, 
we can forget about the statistics 
for the symmetric insulator conditions 
\eqref{LSM_fermion_chain_a}, 
\eqref{LSM_fermion_chain_b}, and \eqref{LSM_fermion_chain_c}. 

\paragraph{Novel Symmetric Insulators:} 
Now, using the above results, we present a few novel symmetric insulator states, 
which are beyond the conventional LSMOH theorem. 
Note that in the conventional LSMOH theorem, usually a single-species 
of charge-1 particles are considered (the important exception to this is the Kondo system, where the mixture of the spin and fermion is considered).  
\begin{itemize}
  \item[(i)]
 We have a trivial symmetric insulator of charge-3 fermion at the filling $\nu =1/3$. This insulator requires the help from the charge-1 trivial fermionic systems. 

The gapping potentials are inherently multi-fermion operators, 
which make the insulator beyond the conventional Slater insulator. 

\item[(ii)]
 We have a trivial symmetric insulator of the mixed system of charge-1 boson at $\nu=1/2$ and charge-1 fermion at $\nu=1/2$. 
This system may be realizable in the optical lattice system.\cite{mix1, mix2}

\item[(iii)]
 We have a trivial symmetric insulator of charge-2 boson at the half-filling. 
This insulator needs the help from the charge-1 fermion systems at the integer fillings. This is related to the superconductors of the fermions, where the Cooper pair of the fermion is bound to the boson and is turned into a neutral boson. 
\end{itemize}

Microscopic lattice Hamiltonians for realizing these insulators 
as well as their higher-dimensional analogues will be left for the future
studies .

\section{$\mathbb{Z}_N$ anomaly and Local Stability}
\label{SPT anomaly and Local Stability}

Having discussed the physics of the chiral anomaly
in relation to the LSMOH theorem,
we now investigate the physics of the first index
in Eq.\ \eqref{two anomaly indices},
the $\mathbb{Z}_N$ anomaly,
assuming 
the continuum theory \eqref{low-energy_Dirac}
is derived as the low-energy effective theory of a (1+1)d lattice fermion model.
This is the part of the full 't Hooft anomaly,
which is unrelated to the LSMOH theorem.
They can be non-zero even when the LSMOH theorem
does not enforce the gaplessness.
In other words, 
the theory with the non-zero first index only
(i.e., with the vanishing chiral anomaly)
can be gapped symmetrically in the lattice system.
This is in sharp contrast with SPT boundaries,
where the both $\mathbb{Z}_N$ and the chiral anomalies
(the full 't Hooft anomaly)
signal the non-perturbative stability. 
The difference can be traced back to
using the effective translation symmetry $G_{eff}= \mathbb{Z}_N$
instead of using the full translation symmetry $G= \mathbb{Z}$.
When $G_{eff}$ is properly extended by using the full symmetry $G$,
then the $\mathbb{Z}_N$ anomaly
can be completely gone and the system can be gapped symmetrically.

This is best illustrated in the following example of the double copies of the low-energy fermions $U(1) \times (G_{eff}= \mathbb{Z}_2)$. 
\begin{align}
H = \int dx\,\sum_{a=1,2} \Psi_a^{\dagger} (-i \partial_x) \sigma^z \Psi_a
\end{align}
with $\mathbb{Z}_2 : \Psi_a \to e^{i\pi(\sigma^z - 1)/2} \Psi_a$.
In terms of Eq.\ \eqref{low-energy_Dirac_Sym}, we have 
$
2\pi \frac{s_R}{N} = \pi
$
and
$\frac{s_L}{N} = 0$.  
Given the natural $(G_{eff} = \mathbb{Z}_2)$-ness of the low-energy symmetry,
we first take $N=2$, i.e., $s_R =1$ and $s_L =0$.
For this state,
the first index is non-zero which the second index (chiral anomaly) vanishes,
i.e.,  
$(\frac{1}{2},0)$.
This non-trivial first index implies that we cannot find the symmetric insulator phase within the $U(1) \times G_{eff}$ symmetry. 

However, the stability enforced by
the $\mathbb{Z}_N$ anomaly index is not non-perturbative in the lattice system
as we saw in the section of the LSMOH theorem:
we can always find a way to gap out the spectrum without breaking symmetries
when the chiral anomaly is absent.
Indeed, if we allow to extend $(G_{eff} = \mathbb{Z}_2 ) \to \mathbb{Z}_4$,
which is still the subgroup in $\mathbb{Z}$,
then $N=4$ with $s_R = 2$ and $s_L =0$.
This generates the completely trivial anomalies, i.e.,
it is labeled by the 't Hooft anomaly $(0,0)$,
and hence it can be gapped without breaking the symmetry
$U(1) \times \mathbb{Z}_4$
(for filling $\nu = \frac{1}{N}$,
by extending the symmetry $\mathbb{Z}_N \to \mathbb{Z}_{N^2}$,
we can remove the $\mathbb{Z}_N$ anomaly completely for any $N$).
Note that we do not allow such extension of the symmetries
at SPT boundaries and hence the $\mathbb{Z}_N$ anomaly
imposes a non-perturbative stability on the edge theory.   

That being said, we may ask what this $\mathbb{Z}_N$ anomaly
means to the low-energy theory of the fermionic lattice system.
The first thing to note is that,
the non-zero $\mathbb{Z}_N$ anomaly implies that the given theory
cannot be gapped within the quadratic term because the theory must be realizable
as the non-trivial SPT boundary.
Even when $\mathbb{Z}_N$ is extended to $\mathbb{Z}$, the quadratic term is not
allowed if the term was originally prohibited by $\mathbb{Z}_N$.
Hence, the non-zero $\mathbb{Z}_N$ anomaly implies that the system is
perturbatively stable.
Furthermore, we can explicitly show that the electronic lattice systems which can be gapped within the quadratic terms do not possess $\mathbb{Z}_N$ anomaly, see Appendix \ref{Proof} for detail. This implies that to gap out the theory with the $\mathbb{Z}_N$ anomaly, we need to include non-perturbative ingredients to the theory, e.g., a help from the extra trivial gapless modes and interactions beyond the terms quadratic in fermions (see, for example, the gapping potentials in the section of LSMOH theorem  \ref{LSMOH_theorem}). Thus the non-zero $\mathbb{Z}_N$ anomaly provides a perturbative stability of the given low-energy fermionic theory. (Note that the vice versa is not true. Even when the $\mathbb{Z}_N$ anomaly is absent, the system may be perturbatively stable.)

\section{(3+1)d Chiral Anomaly and Weyl Semimetals}
\label{(3+1)d Chiral Anomaly and Weyl Semimetals}

Given the relation between
the (1+1)d chiral anomaly and the LSMOH theorem in (1+1)d,
we now ask if the (3+1)d chiral anomaly also contain any stability information.
Here we will show that the (3+1)d (abelian) chiral anomaly provides the local
stability by considering the (3+1)-dimensional relativistic
semimetals\cite{Young2012dirac,yang2015topological,burkov2011weyl,yang2011quantum,cho2011possible}
and chiral anomaly\cite{nielsen1983adler,adler1969axial,bell1969pcac,
  Creutz2001,araki2008non,araki2007anomalies,ishimori2010non} captured by the
triangular
$G$-$U(1)$-$U(1)$ diagram with $G$ being unitary spatial symmetry. 
For example, in the Weyl semimetal, $G$ is the translation symmetry.
$U(1)$ is the external non-dynmical electromagnetic guage field. 

To illustrate that the (3+1)d chiral anomaly does not give non-perturbative
stability related to the filling and the translation,
we take a specific two-band model
with the Bloch Hamiltonian\cite{yang2011quantum} 
\begin{align}
  \mathcal{H}(\bf{k})
  &= \sin(k_x) \sigma^x + \sin(k_y) \sigma^y
    \nonumber \\
  &\quad
    +
    + M (\cos(k_z) - \cos(Q))\sigma^z
    \nonumber\\ 
&\quad  + m (2-\cos(k_x) - \cos(k_y) ) \sigma^z. 
\label{Weyl_Lattice}
\end{align} 
This model has the two Weyl points at $k_z = \pm Q$. Remarkably, the Weyl points
appear at zero energy, which makes the system exactly at half-filling
for any $Q$. Since the system is spinful, the half-filling means that there is
one electron per unit cell, and thus it can be trivially gapped while preserving
the translation and charge conservation symmetries.
Indeed, by changing $Q \to 0$, we can
achieve such a symmetric trivial insulator within
the Bloch Hamiltonian \eqref{Weyl_Lattice} at half-filling.
However, this process involves the change in the dispersion from the relativistic dispersion to the non-relativistic dispersion, which is a non-perturbative process seen from the low-energy Wely fermion Hamiltonian.

On the other hand, in the low-energy limit, the two Weyl points, 
\begin{align}
H = \int d^3x\,\Psi^{\dagger} \tau^z \sigma \cdot k \Psi, 
\end{align}
cannot be gapped while keeping the translation symmetry along $z$   
\begin{align}
T_z = \exp(i Q \tau^z). 
\end{align}
(The other two translation symmetries along $x$ and $y$ directions
are trivial in this fine-tuned model.)
Given the allowed symmetric insulator with translation and the filling,
any stability condition of the Weyl semimetal must be only
local in the parameter space.
This local stability then can be captured by the chiral anomaly $T_z - U(1)- U(1)$ diagram: 
\begin{align}
\delta S =  \int d^4 x ~ \frac{2Q}{16\pi^2} \varepsilon^{\mu\nu\lambda\rho} F_{\mu\nu}F_{\lambda\rho}, 
\label{ChiralAnomaly}
\end{align}
which can be obtained by the change in the path integral measure by the translation symmetry in the presence of the electromagnetic gauge field.

There is another reason why the non-trivial chiral anomaly \eqref{ChiralAnomaly}
may not be equivalent
to the filling-constrained gapless-ness of the original lattice model
\eqref{Weyl_Lattice}.
We essentially show that the number of electrons relevant for the physical
(semi-classical)
picture\cite{parameswaran2014probing,nielsen1983adler,adler1969axial,bell1969pcac}
of the chiral anomaly deviates from that of the original lattice problem.
For this, we count the number of electrons in the system in the presence and in the absence of the mangetic field for the Weyl semimetal. 

\begin{figure}
\begin{center}
\includegraphics[width=\columnwidth]{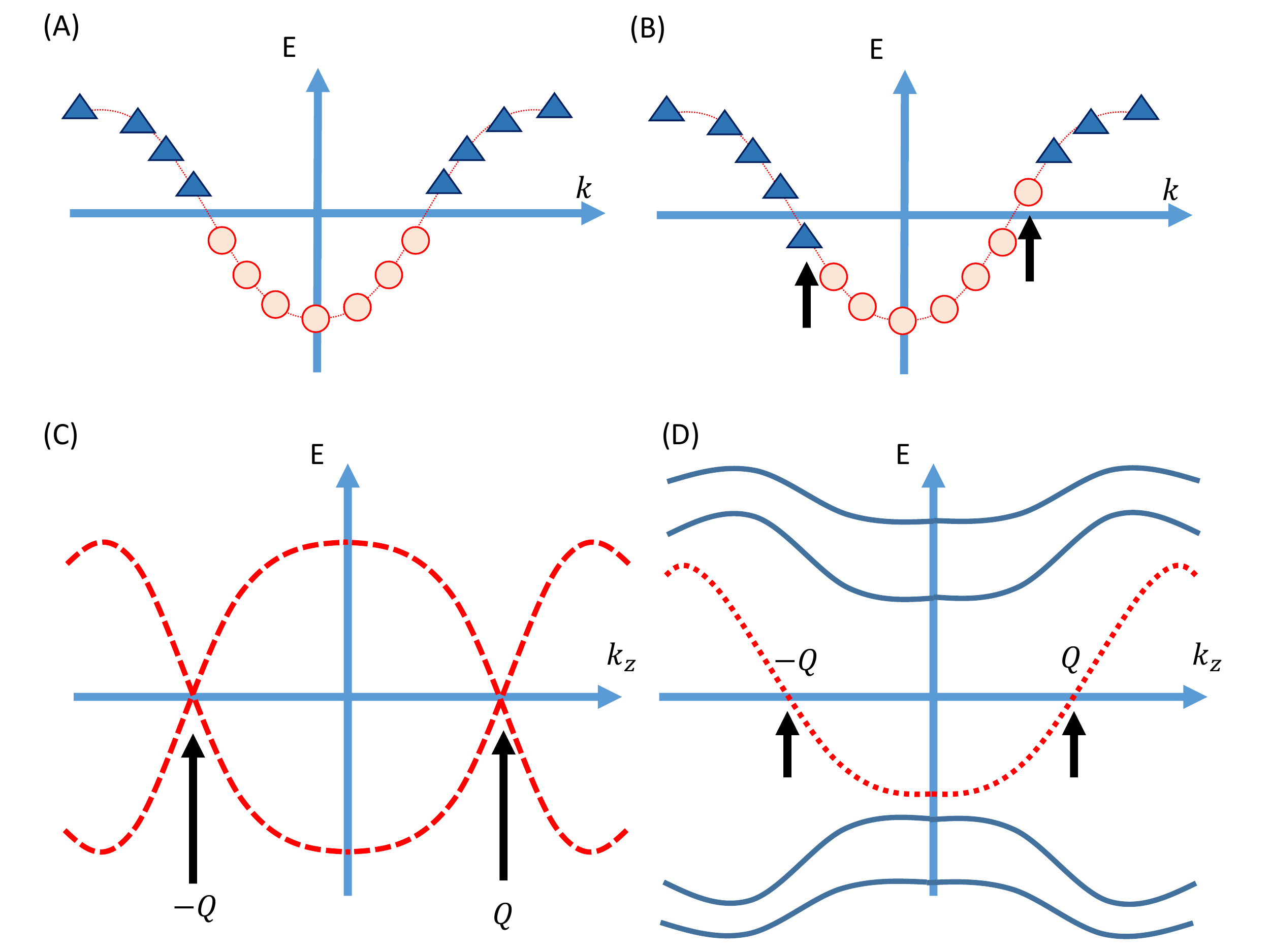}
\caption{Semi-classical illustrations of Anomaly. (A) (1+1)d metallic state. (B) On adiabatic insertion of the flux by $2\pi$, one state at the left is pumped to the right. Equivalently, the momentum labeling each state is shifted by $\frac{2\pi}{L}$. (C) Spectrum of the Wely semimetal in cubic lattice. Number of the state below the chemical potential $\mu=0$ is precisely $L_x \times L_y \times L_z$, which is equivalent to the number of the electrons. (D) On applying the mangetic field, the band structure is changed. }
\label{Fig1}
\end{center}
\end{figure}

First, in the absence of the magnetic field, we note that the spectrum has the
(accidental) particle-hole symmetry, i.e., $E = \pm |E(k)|$.  See (C) of Fig.
\ref{Fig1} for $k_x = k_y = 0$ band structure. We now count the number of the
states below the chemical potential $\mu=0$. Then, there is a single band below
the chemical potential. Hence, there is a single filled state per each momentum
$\vec{k}$. The total number of electrons in the system thus equals to the number
of the allowed momentum. The spacing between the momentum along $a$-direction
($a=x,y,z$) is $\frac{2\pi}{L_a}$ and the each momentum spans from $-\pi$ to $\pi$. Thus, the number of the filled states are: 
\begin{align}
N_{e} = \prod_{a=x,y,z} \Big( \frac{2\pi}{2\pi/L_a} \Big) = L_x L_y L_z, ~ \nu = \frac{N_{e}}{L_x L_y L_z} = 1.\nonumber
\end{align}
Hence there is one electron per unit cell. 
 
Next, if we apply the magnetic field along $z$ direction,
then the band structure is changed into the series of the Landau levels.
At zero chemical potential, there is a chiral mode passing through each Weyl
point.
See Fig.\ \ref{Fig1}(D). There chiral modes are equivalent to the ``0th Landau level" of the Weyl fermions. Now, let us count the number of the filled states in this case. 

We start with counting with the fully filled bands. Given a momentum $k_z$, there are $N_{f-LL} \in \mathbb{Z}$ filled Landau levels with $N_{f-LL}$ varying with the strength of the magnetic field. Then, each $k_z$ has the following number of the filled states 
\begin{align}
N_{e: f-LL} &= \Big( N_{f-LL} \times \frac{L_x L_y}{2\pi l_B^{2}} \Big) \times \frac{2\pi}{2\pi/L_z} \nonumber\\ 
&= N_{f-LL} \frac{L_x L_y L_z}{2\pi l_B^2}. 
\end{align} 
Here $l_B^2 = 1/B$, in which $B$ is the magnetic field strength. Here, to impose
the periodic boundary condition along $x$ and $y$, the number of the states
inside the Landau level, or $\frac{L_x L_y}{2\pi l_B^{2}}$, must be integral,
and hence $N_{e: f-LL}$ is also integral. We next count the number of electrons
in the 0th Landau level. The momentum $k_z$ inside the filled 0th Landau level
expands from $-Q$ to $Q$, (see (D) of Fig.\ \ref{Fig1}) and hence the counting gives 
\begin{align}
N_{e:0-th LL} =  \frac{L_x L_y}{2\pi l_B^{2}}\times \frac{2Q}{2\pi/L_z} = \frac{L_x L_y L_z}{2\pi l_B^2} \frac{2Q}{2\pi}. 
\end{align}

Now we take a filling, which is the number of electrons divided by the volume, 
\begin{align}
\nu = \frac{N_e}{L_x L_y L_z} = \frac{1}{2\pi l_B^2} \Big(N_{f-LL} + \frac{2Q}{2\pi}\Big), 
\end{align} 
where $N_e = N_{e: f-LL}+ N_{e:0-th LL}$.
Obviously, this depends on the separation between the Weyl points $2Q$
and the magnetic field $B$,
and it is not necessarily $\nu =1$.
Hence, the (3+1)d chiral anomaly,
which can be faithfully understood from this Landau level physics,
loses information about the microscopic filling
of the original model,
which is crucial for the existence of a trivial insulator
allowed by the LSMOH theorem.

We now compare this with the 1d chiral anomaly.
In the 1d case, we apply only an electric field adiabatically
to encode the pumping associated with the anomaly
and this effectively is encoded through $k \to k+A(t)$,
in which $A(t)$ is the time-dependent gauge field
varying from $0$ to ${2\pi}/{L}$.
This process does not change the band structure
but only the momentum is shifted.
Hence we shift the state from left end to the right
end as in (A) and (B) of Fig.\ \ref{Fig1} after inserting $2\pi$ flux.
During the process, the filling is not changed and we can directly access
the information of the filling and thus directly to the LSM theorem.

For the Dirac semimetals, e.g., distorted spinel,
Na${}_3$Bi and Cd${}_3$As${}_2$,
\cite{Young2012dirac,yang2014classification, yang2015topological}
where the accidental band crossings are protected by spatial symmetries,
we can also find a similar $G$-$U(1)$-$U(1)$ chiral anomaly.
This can be found in Appendix \ref{Dirac}. 

\section{Conclusions and Outlooks}
\label{Conclusion and Outlooks}

In this paper, we have compared the physics of the LSMOH theorem
and the boundaries of strong SPT phases
from the perspective of quantum anomalies.
We have shown that the same form of the effective theory of the edge of the SPT state can be constructed within the lower-dimensional lattice models. Hence, the no-go theorem for the boundary of the SPTs is circumvented by encoding some on-site symmetry in the strong SPT as the non-on-site translation symmetry in the corresponding lattice model. 

From the connection, we further clarify the implications of the anomalies on the stabilities of the gapless-ness in the two systems. Though the two systems have the identical low-energy theory with the effective symmetry, the anomalies are different in the two systems. The central distinctions between the edge of the SPT and the lattice systems are originated from the non-on-site-ness of the translation symmetry and also from the effective reduction of the translation symmetry. 

By viewing the LSMOH theorem as the anomaly, we have expanded the LSMOH theorem
to the case of the multi-charge and multi-species problems
and constructed several exotic symmetric insulators. 

Finally, we also briefly discussed the (3+1)d chiral anomaly
and have shown that they provide local stability of topological semimetals.

There are several directions to extend the studies here. 

An obvious direction is to include time-reversal symmetry
and other spatial symmetries.\cite{Watanabe2015filling,po2017lattice}
There are several extensions of the LSMOH theorem,
i.e., obstructions to construct a symmetric trivial insulator,
by including time-reversal and several crystalline symmetries.
It would be desirable to interpret these extensions in the language of anomalies. 

Next, given the connection between SPT boundaries and the lattice
systems,
another interesting direction is to clarify, if any,
the distinction between fermionic and bosonic systems in lattice models.
Note that, on SPT boundaries,
fermions and bosons are fundamentally different.
This can be seen from the fact that the SPT classifications of
interacting fermion systems assumes spin structures,
which the bosons are not sensitive to.
Note that, in the lattice systems, we know that spin-statistics connection is
not required,
and thus naively we do not expect to have much distinctions
for the no-go conditions of trivial symmetric insulators
between the fermions and the bosons in the lattice models.
However, from the lights of the physics of SPT phases,
it would be interesting how far the bosons
and fermions are identical or different in the lattice systems.

\acknowledgments 
Authors thank helpful discussion
with Joel Moore, Akira Furusaki, Masaki Oshikawa, Takahiro Morimoto, Eun-Gook Moon, and
Sungbin Lee and acknowledge the financial supports from
the National Science Foundation grant DMR-1455296 (S.R.),
Brain Korea 21 PLUS Project of the Korea government (G.Y.C.)
and NRF of Korea under Grant No. 2016R1A5A1008184 (G.Y.C.).
G.Y.C. also acknowledges the support from Korea Institute for Advanced Study (KIAS) grant funded by the Korea government (MSIP).

\appendix 

\section{Vanishing $\mathbb{Z}_N$ anomaly in Perturbatively-gappable Electronic Systems}\label{Proof}

Here we prove that trivial electronic systems,
which can be perturbatively gapped without breaking
the translation and charge $U(1)$ symmetries,
must have vanishing $\mathbb{Z}_N$ anomaly. 

Imagine that we have $N_f$ species of electrons $a = 1, 2, \cdots N_f$
at filling $\nu_a = \frac{p_a}{n}$ with $p_a \in \mathbb{Z}$
such that $\sum_{a=1 \cdots N_f} p_a =  z \cdot n$
with $z, n \in \mathbb{Z}$ (so that the chiral anomaly vanishes).
Furthermore, by some fine-tuning,
translation symmetry is realized as $\mathbb{Z}_N$ symmetry, so that   
\begin{align}
k_{a,R} = 2\pi \frac{s_{a,R}}{N}, \quad k_{a, L} = 2\pi \frac{s_{a,L}}{N}, 
\end{align}
with $s_{a,R}$ and $s_{a,L}$ taking their values in $\mathbb{Z}$. 

Now when this system can be gapped perturbatively,
i.e., in the quadratic level,
we must have the backscattering term which respects the translation symmetry.
Hence, we can order the momentum in the following way 
\begin{align}
  k_{1,L} = k_{2,R}, \quad
  k_{2,L} = k_{3,R}, \quad \cdots,
  \quad 
  k_{1, R} = k_{N_f, L}. 
\end{align} 
With this, we can now show that 
\begin{align}
  \sum_a \nu_a \frac{(s_{a,R} + s_{a,L})}{\epsilon_N} = 0
  \quad \mod  \mathbb{Z}
\end{align}
To see this, we note that the momenta are labeled as following 
\begin{align}
  &s_{1,R} = \bar{s},
    &
  &s_{1, L} = \bar{s} + \frac{p_1}{n}N,
    \nonumber\\ 
  &s_{2,R} = \bar{s} + \frac{p_1}{n}N,
    &
  &s_{2,L} = \bar{s} + \frac{p_1+p_2}{n}N,
    \nonumber\\ 
  &\cdots~ &&\cdots
             \nonumber\\ 
  &s_{N_f, R} = \bar{s} + \frac{\sum_{a=1}^{N_f -1} p_a}{n}N,
    ~&&s_{N_f, L} = \bar{s} +  \frac{\sum_{a=1}^{N_f} p_a}{n}N
\end{align}
with $p_a \frac{N}{n} \in \mathbb{Z}$ for all $a$ (to keep the translaion symmetry as $\mathbb{Z}_N$) and $\bar{s} \in \mathbb{Z}$.

Now the $\mathbb{Z}_N$ anomaly in this system is 
\begin{align}
\frac{1}{\epsilon_N} (2\bar{s}N_f + z N + 2z N N_f -2 \sum_{a=1}^{N_f} a p_a \frac{N}{n} ) \in \mathbb{Z}, 
\end{align}
where we have used $p_a \frac{N}{n} \in \mathbb{Z}$ and so $2 \sum_{a=1}^{N_f} a p_a \frac{N}{n} \in 2 \mathbb{Z}$. Thus it has vanishing $\mathbb{Z}_N$ anomaly.  

Hence the electronic system which can be perturbatively gapped (equivalently, which can be gapped by the quadratic terms) has a vanishing chiral and $\mathbb{Z}_N$ anomaly.  
  
\section{Dirac Semimetals and Chiral Anomaly}\label{Dirac}

Here we show that the Dirac semimetals
\cite{Young2012dirac,yang2014classification, yang2015topological} 
have the $G$-$U(1)$-$U(1)$ chiral anomaly, which is a manifestation of the local stability given by the spatial symmetries. 

Here we note a few points about the spatial symmetry $G$. In the Dirac semimetal, we distinguish the two symmetry groups: $H$ to realize the relativistic Dirac spectrum in a lattice model, and $G$ for prohibiting the relativistic mass terms to the Dirac spectrum. When seen from the low-energy theory, some elements in $H$ may be superfluous and are not required for the stability. In general, $G$ is a subgroup of $H$. 

Typically, $H$ must contain (i) inversion and (ii) time-reversal symmetries to gaurantee the four-fold degeneracy at the band crossings of Dirac fermions. Furthermore, they accompany (symmorphic or non-symmorphic) rotational symmetries. Otherwise, the dispersions may be gapped or deformed away from the relativistic Dirac spectrum, e.g., line-nodal spectrum. However, as soon as we get to the relativistic spectrum and concentrate only on the relativisitic mass-gap deformation, some of the symmetries in $H$ is not necessary for stabilizing the Dirac semimetal. Hence, $G$ can be smaller than $H$. 

For the known materials of Dirac semimetals, we can show that $G$ can be
generated by only a few orientation-preserving space groups inside $H$, 
and consider $G$-$U(1)$-$U(1)$ anomaly. 

\subsection{Dirac semimetal: disorted spinel}
It has a single Dirac point at the zone boundary. It is an accidental band crossing, not related to the Lieb-Schultz-Mattis theorem. The low-energy Hamiltonian in the chiral basis is given as 
\begin{align}
H = \int d^3k\, \Psi^{\dagger}_{\bm{k}}\tau^z \bm{\sigma}\cdot \bm{k} \Psi_{\bm{k}} 
\end{align}
with two-fold rotation $C_2$ in $xy$-plane, inversion $P$, and time-reversal symmetry $\mathcal{T}$. They are given as following: 
\begin{align}
C_2 = \tau_z \sigma_z, ~ P = \tau_y, ~ \mathcal{T}= i\sigma_y \tau_z \mathcal{K}.  
\end{align}
To keep the relativistic Dirac spectrum, all the three symmetries are required.
However, within the relativistic theory, $C_2 \propto \tau^z$ is enough to
remove the relativistic mass terms.
Obviously, it is captured by the $C_2$-$U(1)$-$U(1)$ chiral anomaly.   
\begin{align}
\delta S =  \int d^4 x ~ \frac{1}{16\pi} \varepsilon^{\mu\nu\lambda\rho} F_{\mu\nu}F_{\lambda\rho} 
\end{align}
We may extend the symmetry group $G$ to be generated by $C_2$ and $P$, which will be isomorphic to $D_8$. 

\subsection{Dirac semimetal: Na${}_3$Bi and Cd${}_3$As${}_2$}
They have two Dirac points on the $k_z$ axis, symmetric under the rotations. They are ``accidental band crossings", not related to the LSM theorem. The low-energy Hamiltonian in the chiral basis is given as 
\begin{align}
H = \int d^3k\, \Psi^{\dagger}_{\bm{k}} \mu^{0} \otimes \sigma^z (\bm{\tau} \cdot \bm{k}) \Psi_{\bm{k}}
\end{align} 
in which $\mu^a$ is the Pauli matrix acting on the ``valley" index. Symmetries are: inversion $P$, time-reversal symmetry $\mathcal{T}$, and 3-fold rotation for Na${}_3$Bi (4-fold rotation for Cd${}_3$As${}_2$). They are given by 
\begin{align}
\mathcal{T} = \mu^x \sigma^z i\tau^y \mathcal{K}, ~ P = \mu^y \sigma^y. 
\end{align}
Translation along $z$-direction is given by 
\begin{align}
T_z = \exp(i Q \mu^z)
\end{align} 
where $(0,0,\pm Q)$ are the positions of the Dirac points.  

The symmorphic 3-fold rotation $C_3$ for Na${}_3$Bi is 
\begin{align}
C_3 = \exp\Big(i\frac{\pi}{3} \sigma_z \otimes \mu^z \Big) \otimes \exp\Big(i\frac{2\pi}{3} \tau^z\Big) 
\end{align}
The symmorphic 4-fold rotation $C_4$ for Cd${}_3$As${}_2$ is 
\begin{align}
C_4 = \mu^z \otimes \sigma^z \otimes \tau^z \exp\Big(-i\frac{\pi}{4}\tau^z\Big). 
\end{align}

The ``stability" statements involve rotation, inversion and time-reversal. In particular, inversion and time-reversal are invoked to gaurantee the four-fold degeneracy at the zero energy (not about the gapless-ness).  

\textit{- Anomaly}:
Now the relativistic gapless-ness is guaranteed
if we impose $T_z$ and $C_n$.
However, $T_z$ and $C_n$ are not anomalous in $g$-$U(1)$-$U(1)$
diagram in which $g$ is generated by composing $T_z$ and $C_n$. 

To see the anomaly structure carefully, we introduce the U(1) valley gauge field $a_\mu$ such that $\mu^z = +1$ fermion carries the charge-$Q$ and $\mu^{z} = -1$ fermion carries the charge-$(-Q)$, i.e., the covariant derivative of the fermions is 
$
D_\mu = \partial_\mu - iA_\mu - i Q \mu^z a_\mu. 
$
Now it is straightforward to compute the triangle diagram in the presence of the
field strength of $A_\mu$ and $a_\mu$,
i.e., $C_n$-$A$-$a$, e.g., for $C_3$ case is 
\begin{align}
\mathcal{L} = \frac{2\pi/3}{16\pi^2} \times Q \times \epsilon^{\mu\nu\lambda\rho} F_{\mu\nu} f_{\lambda\rho}, 
\end{align}
where $f$ is the field strength of $a_\mu$.
Note that $a_\mu$ is the ``gauge field" by gauging on-site version of the
translation.
Hence, this anomaly can be thought of
as $C_3$-``$T_z$"-$U(1)$, where ``$T_z$" is the on-site version of the translation. 

\bibliography{ref}
\end{document}